\documentclass[a4paper,12pt]{article}
\usepackage[utf8]{inputenc}
\usepackage[hidelinks]{hyperref}
\usepackage[margin=2cm]{geometry}
\usepackage{amsmath,amssymb}
\usepackage{authblk}
\usepackage{bm}
\usepackage{graphicx}

\usepackage{lineno}
\usepackage{listings}
\usepackage{tikz}
\usetikzlibrary{arrows}

\usepackage[title]{appendix}

\title{Biological Fluid Mechanics Under the Microscope: A Tribute to John Blake}
\author{David J. Smith\thanks{D.J.Smith@bham.ac.uk}}
\affil{School of Mathematics, University of Birmingham, Edgbaston, Birmingham,
B15 2TT, UK}
\date{}

\begin{document}

\maketitle


\begin{abstract}
  John Blake (1947--2016) was a leader in fluid mechanics, his two principal areas of expertise being biological fluid mechanics on microscopic scales and bubble dynamics. He produced leading research and mentored others in both Australia, his home country, and the UK, his adopted home.
This article reviews John Blake's contributions in biological fluid mechanics, as well as giving the author's personal viewpoint as one of the many graduate students and researchers who benefitted from his supervision, guidance and inspiration. The key topics from biological mechanics discussed are: `squirmer' models of protozoa, the method of images in Stokes flow and the `blakelet' solution, discrete cilia modelling via slender body theory, physiological flows in respiration and reproduction, blinking stokeslets in microorganism feeding, human sperm motility, and embryonic nodal cilia.
\end{abstract}

\section{Introduction}
This tribute to the legacy of the late Professor John Blake (1947--2016) will focus on one of several areas of applied mathematics in which he made major contributions: very low Reynolds number biological fluid mechanics. Including theoretical work which was motivated by biological applications, alongside more directly applied work, John\footnote{John Blake had little interest in formality and in the time I knew him, all of his students and colleagues called him by his first name -- for this reason I will use his first name throughout in this article.} published at least 74 papers in this field between 1971 and 2012, of which twelve at the time of writing (August 2017) have above 100 citations on Google Scholar; four of these landmark papers were published by the time John was 25 years of age. Remarkably, John's output in the distinct area of bubble dynamics was at least as significant -- although I will not address these major achievements in the present article.

A graduate of the University of Adelaide and student of E.O.\ Tuck, John conducted his PhD research with M.J.\ Lighthill in Cambridge, followed by a postdoctoral position at Caltech; he then returned to Australia, working at CSIRO for a period, followed by a Chair at Wollongong, before moving to Birmingham, UK, where he remained for the rest of his career.

Mathematically, the core of the field of very low Reynolds number fluid mechanics is the solution of the Stokes flow equations,
\begin{equation}
-\bm{\nabla} p + \mu\nabla^2 \bm{u} = 0, \quad \bm{\nabla}\cdot \bm{u}=0,
\end{equation}
where \(\bm{u}\) is fluid velocity, \(p\) is pressure and \(\mu\) is dynamic viscosity. These equations describe the nearly inertialess regime of fluid mechanics on microscopic scales -- although they neglect the non-Newtonian effects that may be present, for example in some physiological fluids.

Very low Reynolds number fluid dynamics describes the world encountered by motile protozoa, sperm, bacteria, and the cells making up the epithelia lining the lung, fallopian tubes and the early embryo. While optical and electron microscopes provide us a means of gaining visual information within this realm, intuition about physical principles of propulsion and molecular transport are fraught with difficulty because of the altered balances of inertia versus viscosity, and diffusion versus advection as compared with the everyday world, often leading applied scientists astray. Furthermore, the characteristic geometrical complexity frustrates both standard analytical modelling techniques and even many state of the art computational codes. What follows will describe some of the mathematical tools and the intellectual frameworks that John developed to enable us to understand this world better.

\section{Beginnings: squirming swimmers}
As hinted above, John always encouraged his students to write and promptly submit their work for publication (with varying success!); this approach served him extremely well during his days as a research student. However, his first research project, completed in 1969 as an honours student working with E.O.\ Tuck at the University of Adelaide, was not published until 2010, in a paper dedicated to Tuck \cite{blake2010}. This project concerned the use of the `S-transform',
\begin{equation}
S[f(x)]=\frac{1}{2}\int_{-1}^1 \frac{f(x)-f(t)}{|x-t|} dt, \quad |x|<1,
\end{equation}
which has the Legendre polynomials as eigenfunctions. The transform was applied to approximate very low Reynolds number flow due to moving slender bodies, including sperm-like shapes with a `head' and an elongated `flagellum'. While the mathematical technique was not one which John would use extensively, the application area of biological flow was one in which he would specialise.

Following his Honours' project and having been awarded University of Adelaide George Murray and Commonwealth Scientific and Industrial Research Organisation (CSIRO) scholarships to work in Cambridge, John began reading some relevant literature. One of the papers he worked through was a 1952 study by the famous applied mathematician M.J.\ Lighthill on the squirming motion of a sphere at low Reynolds numbers. Lighthill's paper, which had been published shortly after G.I.\ Taylor's celebrated `swimming sheet' model of propulsion at zero Reynolds number, was part of a series of studies in the early 1950s in which the fundamental question of how microorganisms could propel themselves without the assistance of inertia was solved. As John would point out to his students, these findings were two decades before the celebrated 1976 lecture of E.M.\ Purcell \cite{purcell1977}, `Life at low Reynolds number'. Whereas Taylor studied an infinite swimmer, Lighthill formulated the problem as the solution of the axisymmetric Stokes flow equations around a sphere of radius \(a\) with the `squirming' boundary condition,
\begin{equation}
\begin{aligned}
u_r(a,\theta)      & = \sum_{n=0}^\infty A_n P_n(\cos \theta),\\
u_\theta(a,\theta) & = \sum_{n=0}^\infty B_n V_n(\cos \theta),
\end{aligned}
\end{equation}
where \(P_n\) is the \(n\)th Legendre polynomial, \(V_n=2P_n'(\cos \theta)\sin\theta/(n(n+1))\), and \(A_n\), \(B_n\) are coefficients defining the squirming motion. Lighthill gave a solution for the velocity field \((u_r,u_\theta)\), preceded by the (perhaps dangerous!) phrase \emph{``It is easily seen that the only combination of solutions... satisfying the boundary conditions is given by...''} By seeking solutions only with finite energy, Lighthill was then able to show that the velocity of translation of the sphere is given by \(U=\frac{1}{3}(2B_1-A_1)\).

Lighthill's concise article suppressed a significant amount of calculation, which John carefully attempted to reproduce. He discovered that Lighthill had made an uncharacteristic error -- terms had been omitted from the solutions for both expressions. John's resulting corrected solution took the form,
\begin{equation}
\begin{aligned}
  u_r(r,\theta)       & = -U \cos \theta + A_0 \frac{a^2}{r^2}P_0 + \tfrac{2}{3}(A_1+B_1)\frac{a^3}{r^3}P_1 \\
      & + \sum_{n=2}^\infty \left[\left(\tfrac{1}{2}n\frac{a^n}{r^n}-\left(\tfrac{1}{2}n-1\right)\frac{a^{n+2}}{r^{n+2}}\right)A_n P_n + \left(\frac{a^{n+2}}{r^{n+2}}-\frac{a^n}{r^n}\right)B_n P_n\right],\\
  u_\theta(r,\theta)  & = U \sin \theta + \tfrac{1}{3}(A_1+B_1) \frac{a^3}{r^3} V_1 \\
      & + \sum_{n=2}^\infty \left[\left( \tfrac{1}{2}n\frac{a^{n+2}}{r^{n+2}}-\left(\tfrac{1}{2}n-1\right)\frac{a^n}{r^n} \right)B_n V_n + \tfrac{1}{2}n\left(\tfrac{1}{2}n-1\right)\left(\frac{a^n}{r^n}-\frac{a^{n+2}}{r^{n+2}} \right) A_n V_n \right],
\end{aligned}\label{eq:squirmersol}
\end{equation}
the additional terms being the series with coefficients \(B_n\) in the expression for \(u_r\) and \(A_n\) in the expression for \(u_\theta\).
In John's words,
\begin{quotation}
``This later led to [John Blake] undertaking a Ph.D.\ under the supervision of Sir James Lighthill.. one of the first actions, encouraged by Lighthill, was to publish a corrected version of the paper but with an application directed at ciliary propulsion...'' \cite{smith2012}
\end{quotation}
Motivating the application to ciliary propulsion, John initially focused on the ciliate \emph{Opalina}. Acknowledging that \emph{Opalina} was relatively more disc- than sphere-shaped, he made the following prescient observation,
\begin{quotation}
``...the biological world provides much variety so the following problem may closely model some organism!'' \cite{blake1971spherical}
\end{quotation}
We will return to this observation at the end of the section.

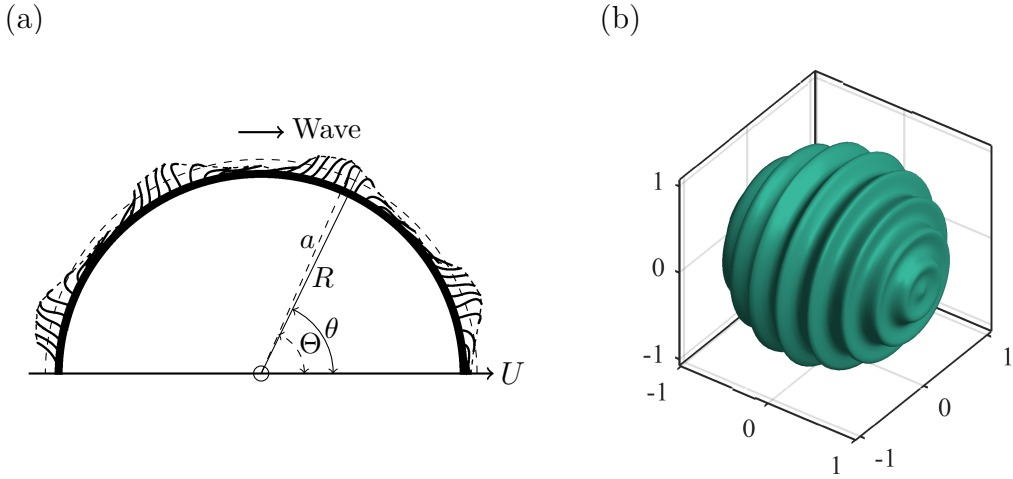
\begin{figure}
    \centering
    \setlength{\tabcolsep}{12pt}
    \begin{tabular}{ll}
      (a) & (b) \\[-0.5em]
      \begin{tikzpicture}
        \node at (0,0) {\includegraphics[trim=0.9cm 0.9cm 0.9cm 0.9cm,clip]{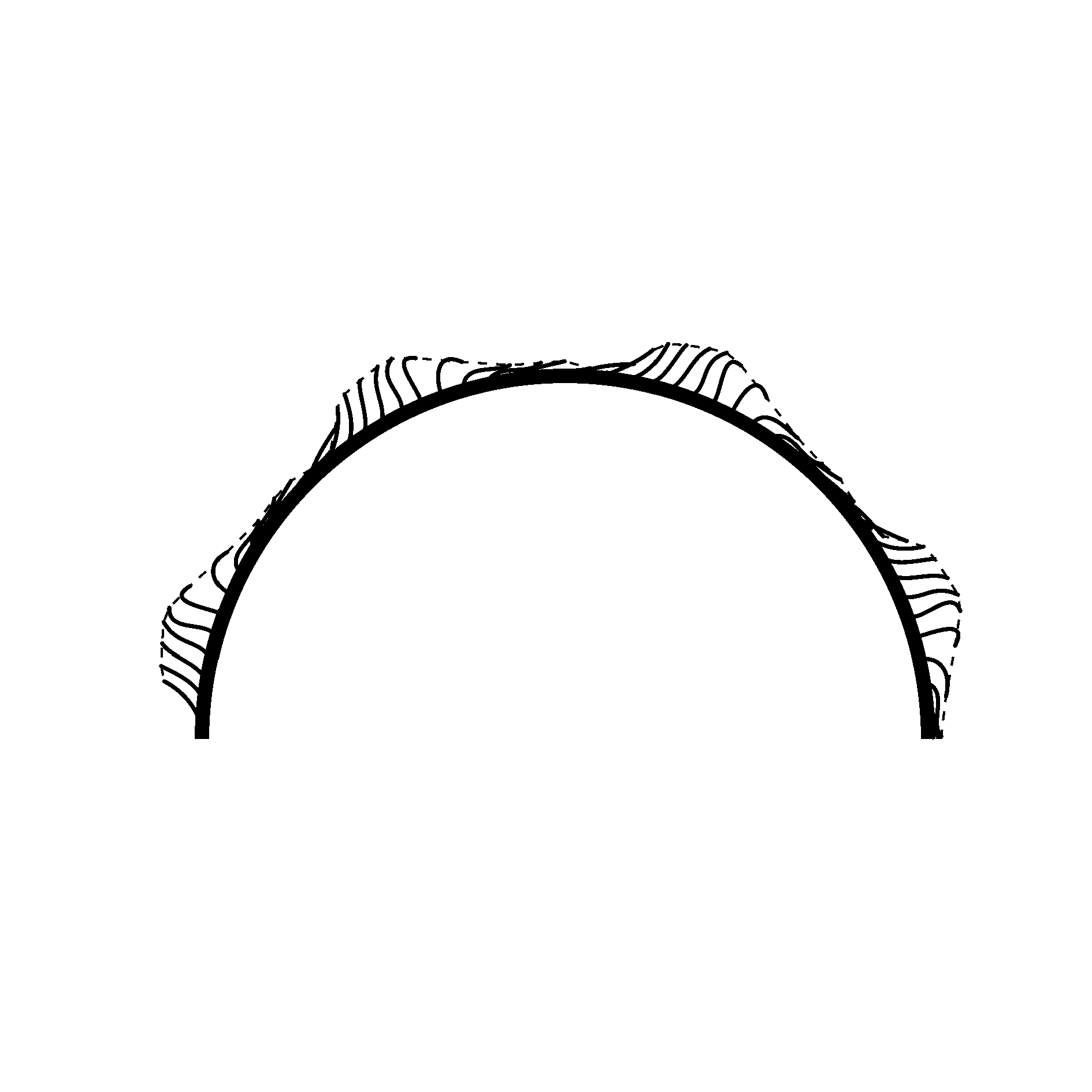}};
        \draw [thick,->] (-2.92,-1.40) -- (3.2,-1.40);
        \draw [dashed] (2.98,-1.40) arc (0:180:2.84);
        \node at (3.45,-1.40) {\(U\)};
        \draw (0.14,-1.40) -- (1.4470,1.3003);
        \draw [dashed] (0.14,-1.40) -- (1.2741,1.2037);
        \node at (0.95,-0.1219) {\small \(R\)};
        \node at (0.75, 0.35) {\small \(a\)};
        \draw (0.14,-1.40) circle (0.1);
        \draw [thick,->] (-0.16,1.80) -- (0.44,1.80);
        \node at (1.00,1.85) {\small Wave};
        \draw [<->] (1.0867,-1.40) arc (0:64.1713:0.9467);
        \draw [<->,dashed] (0.7080,-1.40) arc (0:66.4631:0.5680);
        \node  at (0.79,-0.99) {\small \(\Theta\)};
        \node  at (1.07,-0.79) {\small \(\theta\)};
      \end{tikzpicture}
      &
      \raisebox{0.3cm}{\includegraphics{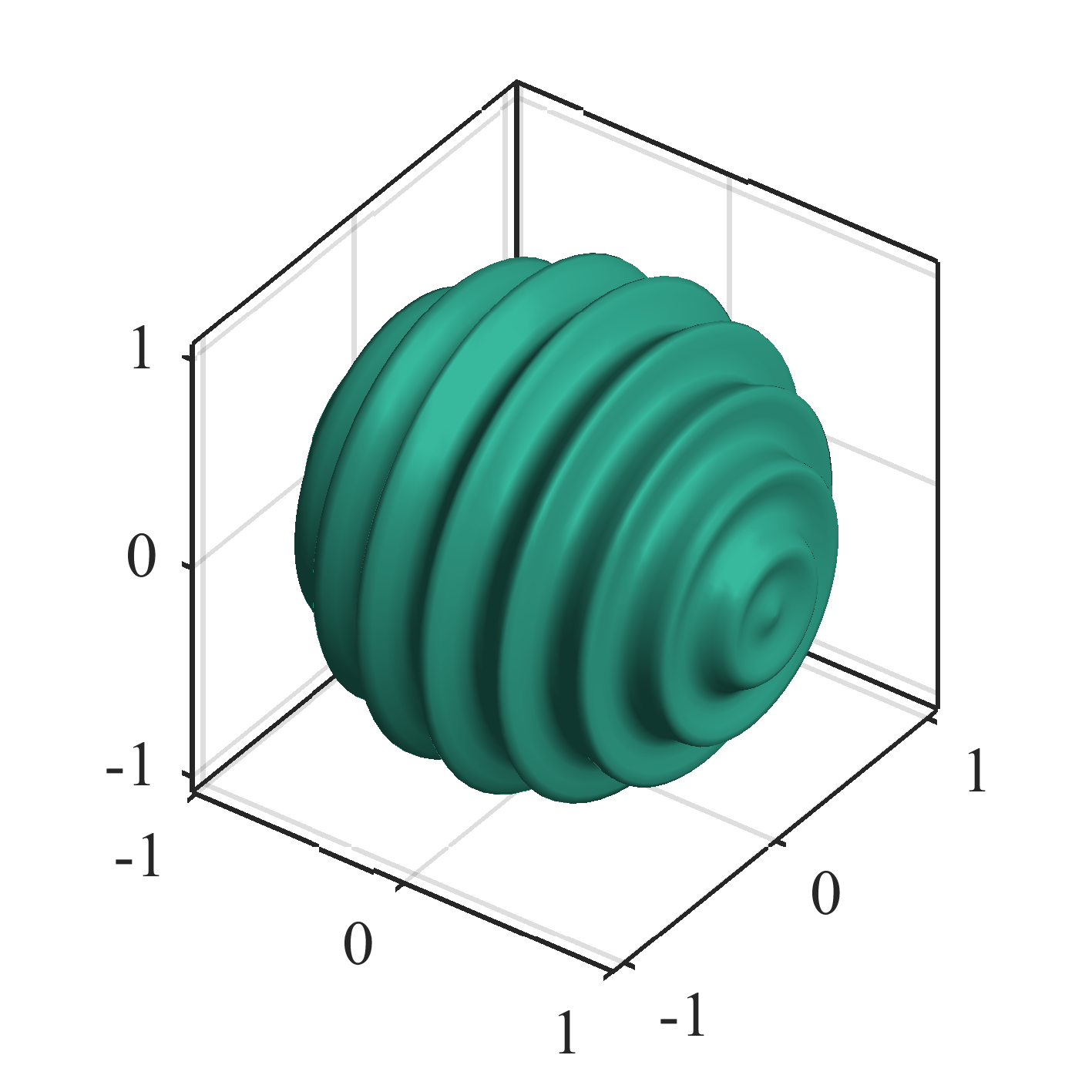}}
    \end{tabular}
    \caption{The squirming envelope model of cilia propulsion, created using models from \cite{blake1971spherical,blake1972model}. (a) Geometric model of a ciliated spherical swimming microorganism, exhibiting the metachronal wave of cilia coordination and `envelope' formed by the cilia tips, from \emph{Opalina} cilia waveforms of \cite{blake1972model} and redrawn as \cite{blake1971spherical}. (b) 3D rendering of a squirmer, using the model of equation~\eqref{eq:squirmerModel} (table 2.4 from ref. \cite{blake1971spherical}).}\label{fig:squirmer}
\end{figure}

Whereas Lighthill had focused on purely tangential motions of the sphere surface, i.e.\ a non-undulating surface, John considered the combination of normal and tangential motions associated with the `envelope' formed by the tips of the densely-spaced cilia (figure~\ref{fig:squirmer}a), which propagates as a \emph{metachronal wave}, typically with cilia synchronised in the direction perpendicular to the direction of propulsion, and phase-shifted proportional to distance along the direction of propagation. A `discrete cilia' model of metachronal coordination is shown in figure~\ref{fig:discreteCiliaBeats}c. Tables of parameters corresponding to various model squirmer motions were constructed, and the associated swimming velocities calculated. One of the models explored is plotted in figure~\ref{fig:squirmer}b; mathematically this model is defined as,
\begin{equation}
  \begin{aligned}
    R(\theta,t)      & = a [  1 + \epsilon((-4.5P_{20}+ 4.5P_{22})\cos(\sigma t) + (-4.4P_{19}+4.4P_{21})\sin(\sigma t))] \\
    \Theta(\theta,t) & = \theta + \epsilon(( 9.3V_{20}-11.7V_{22})\cos(\sigma t) + ( 8.6V_{19}-11 V_{21})\sin(\sigma t)),
  \end{aligned}\label{eq:squirmerModel}
\end{equation}
where \(a\) is equilibrium radius, \(\epsilon\) is amplitude of surface perturbation relative to the radius, and \(\sigma\) is radian beat frequency.

The coefficients \(A_n\), \(B_n\) in the solution~\eqref{eq:squirmersol} were then found by expressing the boundary shape, and hence velocity, in terms of spherical harmonics, approximating through a small-amplitude expansion about the undeformed sphere surface, and applying the no-slip, no-penetration condition. As Taylor had previously found in his analysis of a swimming infinite sheet, the propulsive velocity \(\bar{U}\) was found at second order in the expansion, i.e.\ proportional to the square of deformation height. After calculation of the rate of working \(P\), the `hydrodynamical efficiency' \(\eta=6\pi\mu a \bar{U}^2/\bar{P}\) could then be derived. With parameters \(a=100\)~\(\mu\)m, \(\sigma=25\)~s\(^{-1}\) and \(\epsilon=0.05\), the propulsive velocity was calculated to be \(104.8\)~\(\mu\)m/s, comparable to the swimming velocities observed in nature.

John explored other similar models shortly afterwards, including planar and cylindrical infinite swimmers \cite{blake1971infinite,blake1971self}, however he predominantly focused on discrete-cilia modelling and boundary integral/singularity-based methods throughout most of his career. Nevertheless, the approach he developed with Lighthill was taken up by many other researchers: a Google Scholar search on 14th August 2017 brought up 417 citations. Recent studies inspired by the Lighthill--Blake squirmer include analysis in unsteady Stokes flow \cite{ishimoto2013}, nonlinear dynamics of swimmers in Poiseuille flow \cite{zottl2012}, and most recently, the multicellular algae \emph{Volvox} \cite{pedley2016}. The latter paper returned `full circle' to his lifelong friend and the examiner of his PhD thesis T.\ J.\ Pedley FRS, with R.\ Goldstein FRS and D.\ Brumley. \emph{Volvox} is perhaps the perfect example of a squirming sphere that John had, in 1971, anticipated would be waiting in nature.

\section{Slender body theory and the blakelet} \label{sec:disccilia}
The squirming envelope model of ciliary propulsion, while analytically tractable and informative about species in which the tips of the cilia remain closely-spaced, is nevertheless limited in explaining some of the details of how the cilia beat pattern, and interaction with the cell body, affect propulsion and flow. The model also fails to describe widely-spaced cilia such as those in the embryonic node. John's aim therefore was to model an individual cilium, and then to exploit periodicity to scale up to a dense, coordinated field. Cilia, like the flagella of sperm, are slender organelles, therefore he sought to extend the slender body theory approach pioneered for sperm propulsion by Lighthill's student G.J.\ Hancock \cite{hancock1953,gray1955} to the situation of a slender object projecting from a much larger body, as is the case for ciliated protozoa.

The key idea of slender body theory is to exploit the linearity of the Stokes flow equations and model the object by a line distribution of fundamental solutions, or \emph{stokeslets}, to use the term coined by Hancock. Consider the solution \((\bm{u},p)\) to the singularly-forced Stokes flow equations,
\begin{equation}
  0 = -\bm{\nabla}p+\mu\nabla^2 \bm{u} + \bm{F}\delta(\bm{x}-\bm{\xi}), \quad \nabla \cdot \bm{u}=0.
\end{equation}
In three dimensions and with boundary condition \(\bm{u}\rightarrow 0\) as \(|\bm{x}|\rightarrow \infty\), the velocity field can be expressed as, \(u_j(\bm{x})=F_kS_{jk}(\bm{x},\bm{\xi})\) where \(S_{jk}\) is the second-rank tensor known as the stokeslet (here and in what follows, the summation convention will be used),
\begin{equation}
S_{jk}(\bm{x},\bm{\xi}) = \frac{1}{8\pi\mu}\left(\frac{\delta_{jk}}{r}+\frac{r_j}{r_k}{r^3}\right),
\end{equation}
where \(r_j=x_j-\xi_j\) and \(r^2=r_jr_j\).

\subsection{The image system for a stokeslet near a plane boundary}
Boundaries, such as the body of a protozoa, ciliated epithelium, or glassware in the laboratory, have a critical long-range effect in Stokes flow. To take this effect into account, the most basic problem to solve is to determine how the flow field produced by a concentrated force is changed by the presence of an infinite plane boundary, represented mathematically by the boundary condition \(\bm{u}(x_1,x_2,0)=0\) -- indeed because the cilium is much shorter than the radius of curvature of a typical ciliated microorganism, the ciliated surface can be modelled as an infinite plane \(x_3=0\). I recall John telling the story of how he found various texts in which the authors stated that the solution for the Stokes flow given by a point force in the vicinity of a plane boundary was well-known; however on writing to the authors he found that none were able to tell him exactly where the solution had been written down! He therefore sought to find a solution himself, using the method of images -- i.e.\ for a singularity located at \(\bm{\xi}=[\xi_1,\xi_2,h]\), a fictitious `image' would be placed outside the flow domain at \(\bm{\xi}^*=[\xi_1,\xi_2,-h]\), the problem was to find the form of this image. This task is substantially more difficult than the familiar example in potential theory with a Dirichlet condition because velocity is a vector rather than a scalar. An image stokeslet alone cannot cancel components of the flow both tangential and normal to the surface. 

\begin{figure}
  \centering
  \begin{tabular}{l}
    (a)
    \\
  \begin{tikzpicture}
    \draw [-latex] (0,0) -- ( 9.00, 0.00);
    \draw [-latex] (0,0) -- ( 0.00, 3.67);
    \draw [-latex] (0,0) -- (-1.67, 1.17);
    \node at ( 8.90,-0.30) {\small \(x_1\)};
    \node at ( 0.32, 3.50) {\small \(x_3\)};
    \node at (-1.35, 1.30) {\small \(x_2\)};
    \draw [-latex,thick] (1.67, 1.33) -- (3.33, 1.33);
    \draw [-latex,thick] (3.33,-1.33) -- (1.67,-1.33);
    \draw [dashed] (0,0) -- (2.5, 1.33);
    \node at (1.05, 0.81) {\(\bm{\xi} \)};
    \draw [dashed] (0,0) -- (2.5,-1.33);
    \node at (1.00,-0.85) {\(\bm{\xi}^*\)};
    \node at (2.5, 1.70) {\small stokeslet (\( F\))};
    \node [anchor=north west] at (-1.97,-1.60) {
      \small \begin{tabular}{cccc}
          \begin{tabular}{c} Image system: \\ \quad    \\ \quad \end{tabular}
          &
          \begin{tabular}{c} stokeslet     \\ (\(-F\)) \\ \quad \end{tabular}
          &
          \begin{tabular}{c} stokes-       \\ doublet  \\ (\(2hF\)) \end{tabular}
          &
          \begin{tabular}{c} source-       \\ doublet  \\ (\(-4\mu h^2 F\)) \end{tabular}
        \end{tabular}
      };
    \node at (3.5, 3.40) {\small far-field: \quad stokes-doublet};
    \draw [-latex,thick] (7.03,3.13) -- (7.03,3.73);
    \draw [-latex,thick] (6.38,3.68) -- (6.38,3.08);
    \filldraw (6.70,3.40) circle (0.05);
    \draw [-latex,thick] (6.42,3.73) -- (7.03,3.73);
    \draw [-latex,thick] (6.98,3.08) -- (6.38,3.08);
    \draw [-latex,thick] (5.05,-1.63) -- (5.05,-0.98);
    \draw [-latex,thick] (4.40,-1.03) -- (4.40,-1.68);
    \filldraw (4.725,-1.33) circle (0.05);
    \filldraw (6.850,-1.33) circle (0.05);
    \filldraw (7.180,-1.33) circle (0.05);
    \node at (3.850,-1.33) {\(+\)};
    \node at (5.900,-1.33) {\(+\)};
  \end{tikzpicture}
  \\
  (b)
  \\
  \begin{tikzpicture}
    \draw [-latex] (0,0) -- ( 9.00, 0.00);
    \draw [-latex] (0,0) -- ( 0.00, 3.67);
    \draw [-latex] (0,0) -- (-1.67, 1.17);
    \node at ( 8.90,-0.30) {\small \(x_1\)};
    \node at ( 0.32, 3.50) {\small \(x_3\)};
    \node at (-1.35, 1.30) {\small \(x_2\)};
    \draw [-latex,thick] (2.50, 0.50) -- (2.5, 2.167);
    \draw [-latex,thick] (2.50,-0.50) -- (2.5,-2.167);
    \draw [dashed] (0,0) -- (2.5, 1.33);
    \node at (1.05, 0.81) {\(\bm{\xi}  \)};
    \draw [dashed] (0,0) -- (2.5,-1.33);
    \node at (1.00,-0.85) {\(\bm{\xi}^*\)};
    \node at (3.85, 1.33) {\small stokeslet (\( F\))};
    \node [anchor=north west] at (-1.97,-2.10) {
      \small \begin{tabular}{cccc}
          \begin{tabular}{c} Image system: \\ \quad    \\ \quad \end{tabular}
          &
          \begin{tabular}{c} stokeslet     \\ (\(-F\)) \\ \quad \end{tabular}
          &
          \begin{tabular}{c} stokes-       \\ doublet  \\ (\(-2hF\)) \end{tabular}
          &
          \begin{tabular}{c} source-       \\ doublet  \\ (\(4\mu h^2 F\)) \end{tabular}
        \end{tabular}
      };
    \node at (5.0, 3.10) {\small \begin{tabular}{cc} \begin{tabular}{l} far-field:\\ \quad \end{tabular} & \begin{tabular}{l}stokes-quadrupole\\ source-doublet \end{tabular} \end{tabular}};
    \draw [-latex,thick] (4.725,-1.21) -- (4.725,-0.61);
    \draw [-latex,thick] (4.725,-1.45) -- (4.725,-2.05);
    \filldraw (4.725,-1.33) circle (0.05);
    \filldraw (6.965,-1.50) circle (0.05);
    \filldraw (6.965,-1.17) circle (0.05);
    \node at (3.6125,-1.33) {\(+\)};
    \node at (5.8450,-1.33) {\(+\)};
  \end{tikzpicture}
  \\
    \begin{tabular}{ll}
      (c)
      &
      (d)
      \\
      \includegraphics[width=6.5cm]{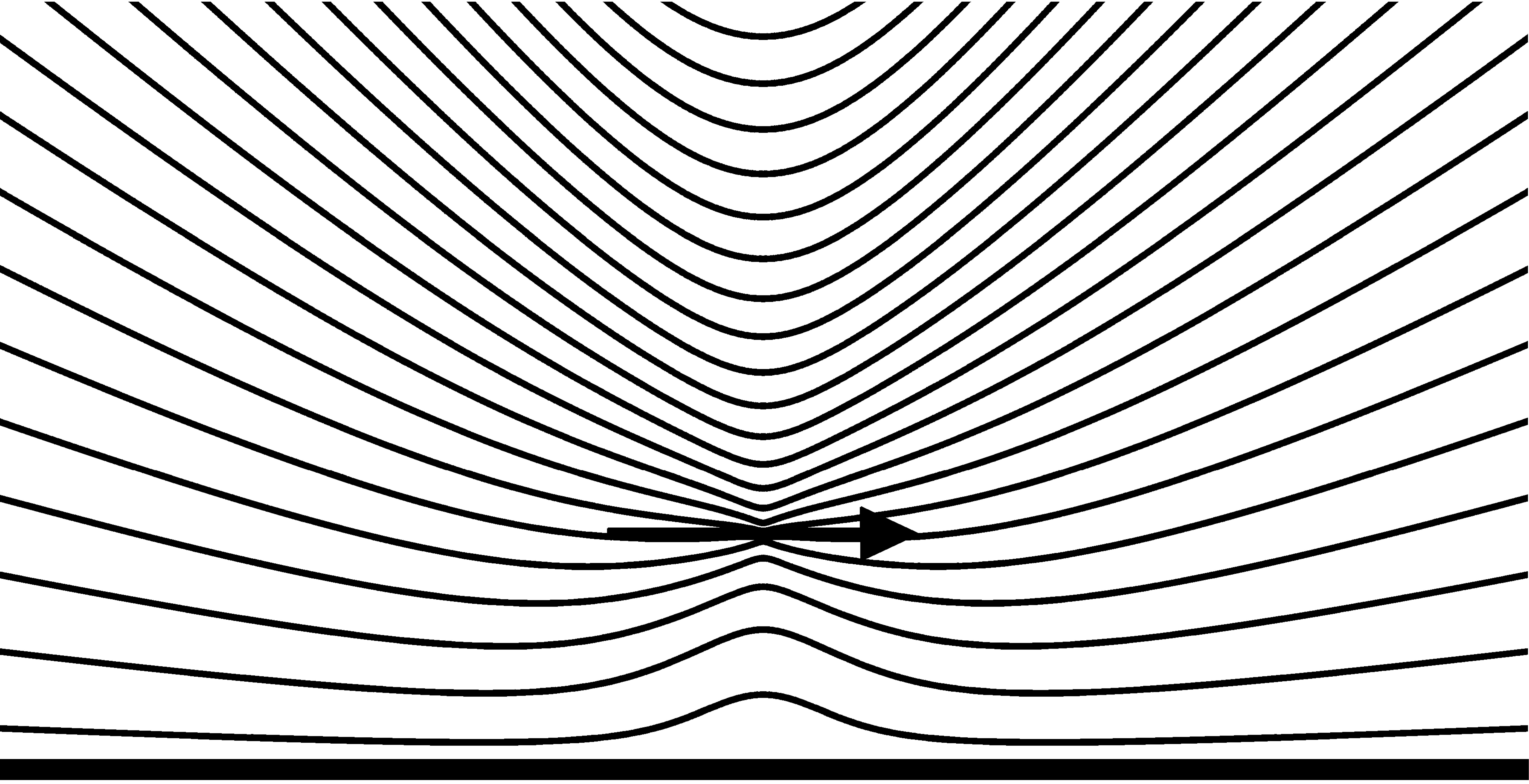}
      &
      \includegraphics[width=6.5cm]{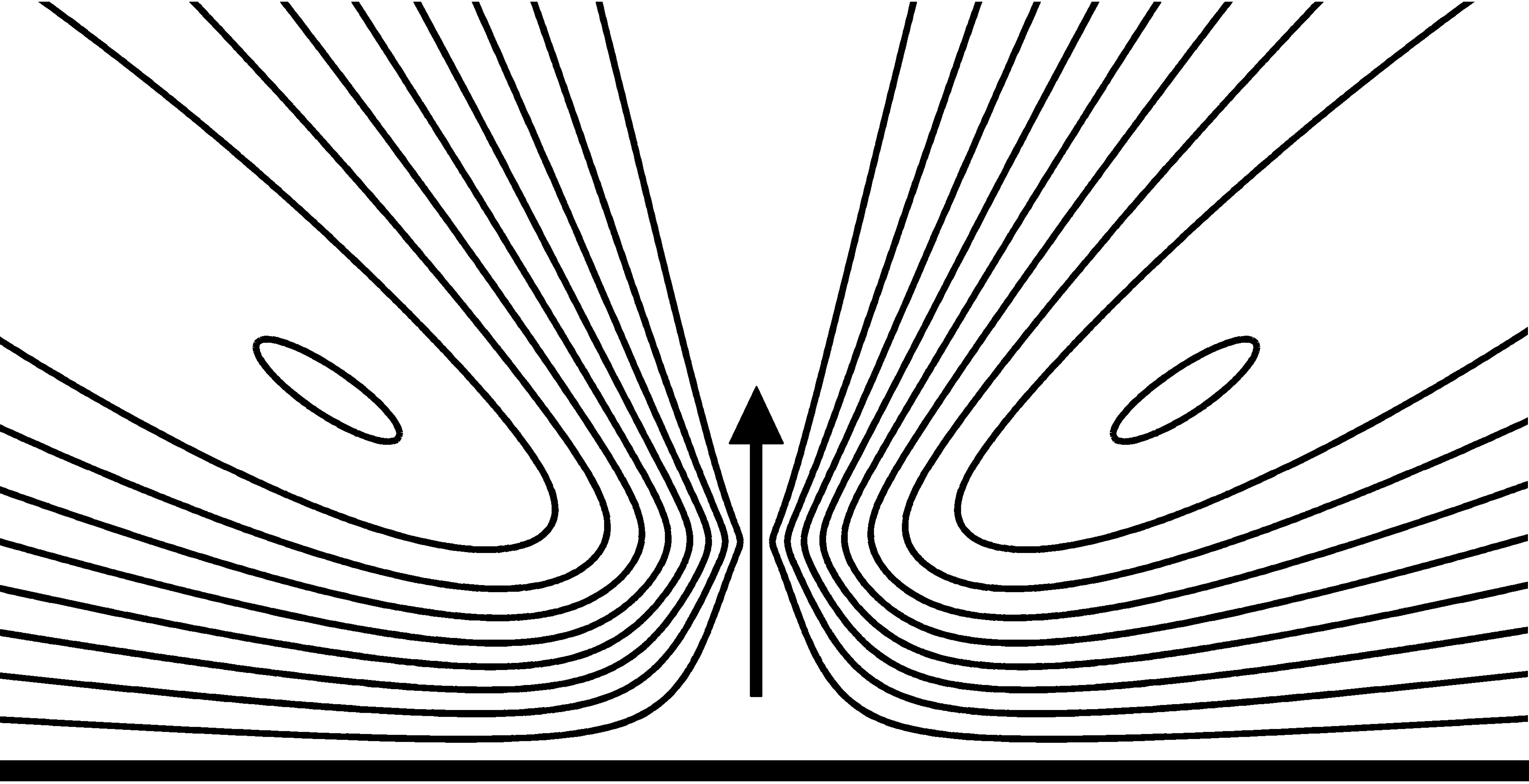}
    \end{tabular}
  \end{tabular}
  \caption{The image systems and streamlines of a stokeslet a distance \(h\) from a solid plane boundary at \(x_3=0\). (a, b) images for parallel and perpendicular oriented stokeslets, redrawn from ref.~\cite{blake1971note}. (c, d) Streamlines for parallel and perpendicular stokeslets respectively; note the radial streamlines in the far-field of a parallel stokeslet, and the closed streamlines associated with a perpendicular stokeslet.}\label{fig:images}
\end{figure}
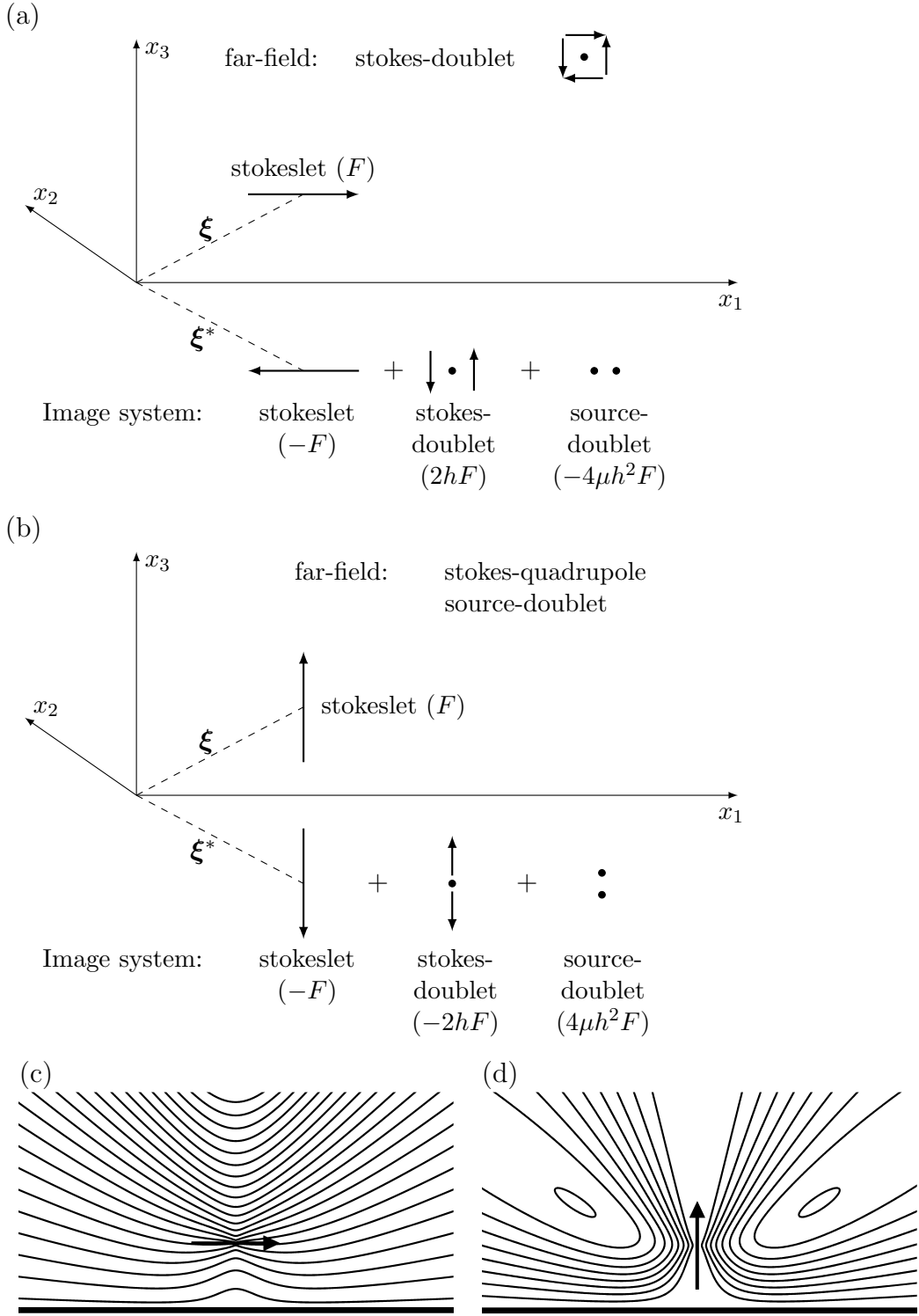

John's approach, inspired by elasticity theory, was to make use of a two-dimensional Fourier transform to find the correction to the above naive images. Having solved the problem and inverted the Fourier transform, the solution \(G_{jk}\), given in ref.~\cite{blake1971note}, can be expressed as,
\begin{equation}
  \begin{aligned}
    u_j(\bm{x}) & = F_k G_{jk}(\bm{x},\bm{\xi}) \\
    & = \frac{F_k}{8\pi\mu} \left[\left\{\frac{\delta_{jk}}{r}+\frac{r_jr_k}{r^3}\right\}-\left\{\frac{\delta_{jk}}{R}+\frac{R_jR_k}{R^3}\right\}\right.
      \\
      & \left. + 2h(\delta_{k\alpha}\delta_{\alpha l}-\delta_{k3}\delta_{3l})\frac{\partial}{\partial R_l} \left\{\frac{h R_j}{R^3} - \left(\frac{\delta_{j3}}{R}+\frac{R_jR_3}{R^3}\right) \right\}\right]
  \end{aligned}
\end{equation}
where \(\alpha=1\) or \(2\), and \(r\) and \(R\) are defined as follows,
\begin{equation}
  \begin{aligned}
    r & = [(x_1-\xi_1)^2+(x_2-\xi_2)^2+(x_3-h)^2]^{\frac{1}{2}} \\
    \mbox{and} \quad \quad
    R & = [(x_1-\xi_1)^2+(x_2-\xi_2)^2+(x_3+h)^2]^{\frac{1}{2}},
  \end{aligned}
\end{equation}
the variables \(r_i\) and \(R_i\) being apparent from these definitions. The term \((\delta_{k\alpha}\delta_{\alpha l}-\delta_{k3}\delta_{3l})\) is simply \(+1\) if \(j=k=1,2\) and \(-1\) if \(j=k=3\); the derivative term on the second line corresponds to a source-dipole and a stokes-doublet (i.e.\ the first derivative of the stokeslet). A schematic representation of the image systems and a plot of their streamlines is given in figure~\ref{fig:images}.

John acknowledged in his manuscript that the solution was, in principle, available from the reciprocal theorem of H.A.\ Lorentz. However, as John observed, \emph{``...the present method yields much more clearly the form of the image system.''} For this reason, the solution is sometimes referred to as the \emph{blakelet} in his honour.

With knowledge of the image system structure, the far-field behaviour could then be inferred, providing insight into how cilia induce fluid transport. A stokeslet oriented parallel to a boundary produces a stokes-doublet far-field, also known as a `stresslet', decaying as \(r^{-2}\), whereas a stokeslet oriented perpendicular to a boundary produces a stokes-quadrupole/source-doublet far-field, decaying as \(r^{-3}\). This form of the far-field was described in ref.~\cite{blake1971note} and given explicitly in ref.~\cite{blake1974fundamental} through the formula,
\begin{equation}
  \begin{aligned}
    u_j(\bm{x}) \sim \frac{F_k}{8\pi\mu} \left[\frac{12 h x_j x_\alpha x_3 \delta_{k\alpha}}{|\bm{x}|^5} + h^2 \delta_{k3} \left(-\frac{(12+6\delta_{j3})x_jx_3}{|\bm{x}|^5}+\frac{30x_j x_3^3}{|\bm{x}|^7} \right) \right].
  \end{aligned}
\end{equation}

This expression also leads to the volume flow rate produced by a stokeslet parallel to a boundary, which in fact was only explicitly stated in a paper by N.\ Liron \cite{liron1978},
\begin{equation}
q_1=\int_{-\infty}^{\infty} \int_{0}^{\infty} F_1 G_{11}(x_1,x_2,x_3;\xi_1,\xi_3,h) dx_2 dx_3 \rightarrow \frac{F_1h}{\pi\mu} \quad \mbox{as} \quad x_1\rightarrow \infty.
\end{equation}
As will be described later, this expression is valuable in assessing the propulsive effectiveness of various cilia motions.

An additional benefit of John's methodology was that the Fourier transform technique could be applied to other problems. As a postdoc at Caltech and in collaboration with A.T.\ Chwang, he derived analogous results for a point-torque singularity (`rotlet'), source and source-doublet, in the vicinity of a solid plane boundary \cite{blake1974fundamental}. Another important example is that for a point force between two parallel plates, representing the environment between a microscope slide and coverglass. I recall John stating that he had solved this problem as far as finding a solution in Fourier space but did not publish it; N.\ Liron \& S.\ Mochon \cite{liron1976} subsequently used his method and expressed the solution in terms of both Hankel integrals and infinite series. A subsequent study has expressed this solution in terms of two Blake image systems and a rapidly-decaying Hankel integral which is ideally-suited for numerical use \cite{staben2003}. These solutions are of such value that even recently, modified expressions, regularized versions and doubly-periodic versions have been developed (\cite{yeomans2016,cortez2015,hoffmann2017}).

The existence of eddies produced purely by the presence of boundaries in viscous flow was first evident from John's derivation of the the image system for a stokeslet acting perpendicular to a plane wall (figure~\ref{fig:images}d). Liron, building on John's Fourier transform technique, found that a stokeslet acting between parallel plane solid boundaries would also produce eddies \cite{liron1978}, a finding which is generically relevant to the flow patterns produced by motile microorganisms viewed in a typical microscopy set up. Further studies of eddies produced in cylindrical geometries, and by ring distributions of stokeslets near plane boundaries followed; a comprehensive collection of these results was given by Liron \& Blake \cite{liron1981}. Their overall summary of this research was that, \emph{``...flow fields produced by sessile micro-organisms are determined primarily by the container geometry in which they are located...''} a crucial finding for any microscopist studying such species to be aware of.

\subsection{Discrete cilia modelling}

The construction of the plane boundary image system then enabled John to pursue a slender body theory based on line distributions of these singular solutions. While the majority of work on slender body theory has sought high precision -- for example through higher-order corrections -- John reasoned that this level of mathematical precision was unwarranted for the biopropulsion problem, and unlikely to result in significantly different predictions. The model he developed would be the first to take into account detailed movement of the cilium beat, and the associated wall-interaction and cilium orientation effects that produce qualitatively different results from the surface squirmer model. It would also form the basis for later studies of physiological cilia-driven flow.

\begin{figure}
  \centering
  \setlength{\tabcolsep}{12pt}
  \begin{tabular}{ll}
    (a)
    &
    (b)
    \\
  \begin{tikzpicture}
    \draw [->] (0,0) -- (4,0);
    \draw [->] (0,0) -- (0,4);
    \draw [->] (0,0) -- (-1.5,2);
    \draw (0,0) circle (0.05);
    \node at (3.9,-0.3) {\(X_1\)};
    \node at (0.0,-0.3) {\(O\)};
    \node at (0.3, 3.8) {\(X_3\)};
    \node at (-1.5, 2.2) {\(X_2\)};
    \node at (-0.70,0.10) {\small \(s=0\)};
    \node at (3.90, 3.60) {\small \(s=L\)};
    \draw [dashed] plot [smooth,tension=0.5] coordinates {( 0.00,0) ( 0.05,0.5) (0.33,1.18) (0.70,1.70) (1.50,2.50) (2.40,3.10) (3.30,3.50)};
    \draw [thick] plot [smooth,tension=0.5] coordinates    {(-0.14,0) (-0.09,0.6) (0.21,1.28) (0.56,1.77) (1.38,2.61) (2.34,3.23) (3.22,3.58) (3.30,3.38) (2.46,2.97) (1.62,2.39) (0.84,1.62) (0.45,1.08) (0.16,0.45) (0.14,0)};
    \draw (0,0) -- (2.40,3.10);
    \node at (2.05,1.75) {\(\bm{\xi}(s,t)\)};
  \end{tikzpicture}
  &
  \begin{tikzpicture}
    \draw [-latex] (0,0) -- (3.2,4.0);
    \draw [-latex] (0.8,1.0) -- (-1.2,2.6);
    \draw [thick]  (0.2,-0.16) -- (2.6,2.84);
    \draw [thick]  (-0.2,0.16) -- (2.2,3.16);
    \node at (4.05,3.7) {\(\bm{t}\), \(V_T\), \(F_T\)};
    \node at (-0.6,2.9) {\(\bm{n}\), \(V_N\), \(F_N\)};
    \draw plot [smooth,tension=1] coordinates {(0.2,-0.16) (0.08,0.1) (-0.2,0.16)};
    \draw plot [smooth,tension=1] coordinates {(0.2,-0.16) (-0.08,-0.1) (-0.2,0.16)};
    \draw plot [smooth,tension=1] coordinates {(2.6,2.84) (2.48,3.1) (2.2,3.16)};
  \end{tikzpicture}
  \end{tabular}
  \caption{The coordinate systems of the discrete cilia model, redrawn from ref.~\cite{blake1972model}. (a) The global coordinate system, where \(s\) is arclength, \(t\) is time and \(\bm{\xi}(s,t)\) is the curved centreline. (b) The local coordinate system, where \(\bm{n}\) and \(\bm{t}\) are normal and tangent vectors, \(V_N\) and \(V_T\) are normal and tangential velocity components, and \(F_N\) and \(F_T\) are normal and tangential components of force per unit length exerted by the cilium on the fluid.}\label{fig:disccilia}
\end{figure}
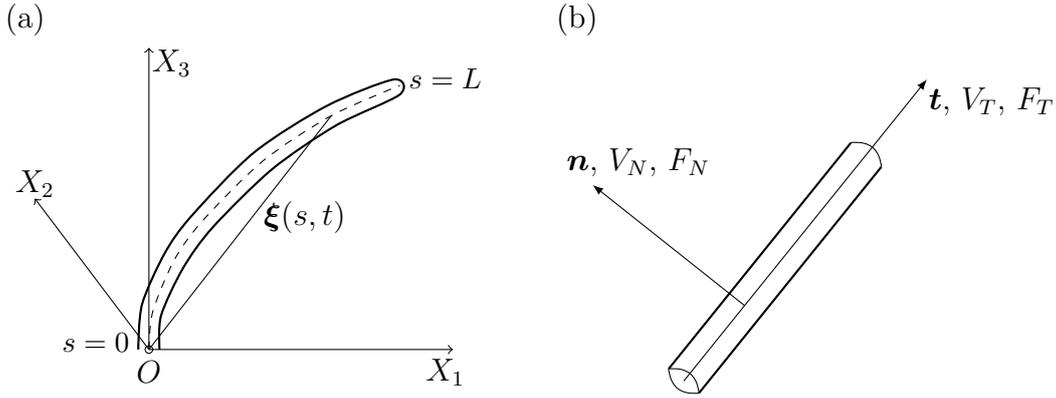

In this section we will outline the mathematical model developed in ref.\ \cite{blake1972model}, John's seminal first treatment of this formidable fluid mechanics problem. The first step in formulating such a model is to consider the flow produced by an individual cilium. The set-up is shown in figure~\ref{fig:disccilia}a: a cilium has instantaneous shape given as a function \(\bm{\xi}\) of arclength \(s\in[0,L]\) measured from the base at the origin \(\bm{X}=O\), and time \(t\). The force per unit length exerted by the cilium on the fluid is denoted \(\bm{F}(\bm{\xi})\), so that the velocity field produced by the cilium is of the form of a line integral of stokeslets and their images,
\begin{equation}
  u_j(\bm{x},t) = \int_0^L F_k[\bm{\xi}(s,t)] G_{jk}(\bm{x},\bm{\xi}(s,t)) ds.
\end{equation}
The analysis of the far-field of the image system then yields the far-field form,
\begin{equation}
  u_j\sim \frac{3r_3^*}{2\pi\mu} \int_0^L \frac{r_j^* r_\alpha^*}{r^{*5}} \xi_3 F_\alpha ds + O\left(\frac{1}{r^{*3}}\right), \label{eq:singlecilium}
\end{equation}
where \(r_j^*=\xi_j-\xi_j^*\).

Cilia beat forwards and backwards to produce flow. The time-invariance of the Stokes flow equations means that the forward and backward strokes must be different to enable an overall flow to be produced. During the forward or `effective' stroke, the cilium is relatively extended, while during the backward or `recovery' stroke, the cilium moves much closer to the surface. John noted from equation~\eqref{eq:singlecilium} that the region around the force singularity which was effectively propelled was proportional to \(\xi_3\). By extending further from the surface during the effective stroke, more fluid would be propelled than during the recovery stroke, producing net transport.

\begin{figure}
  \begin{tabular}{l}
    \begin{tabular}{ll}
      (a) & (b)
      \\
      \includegraphics[width=7cm]{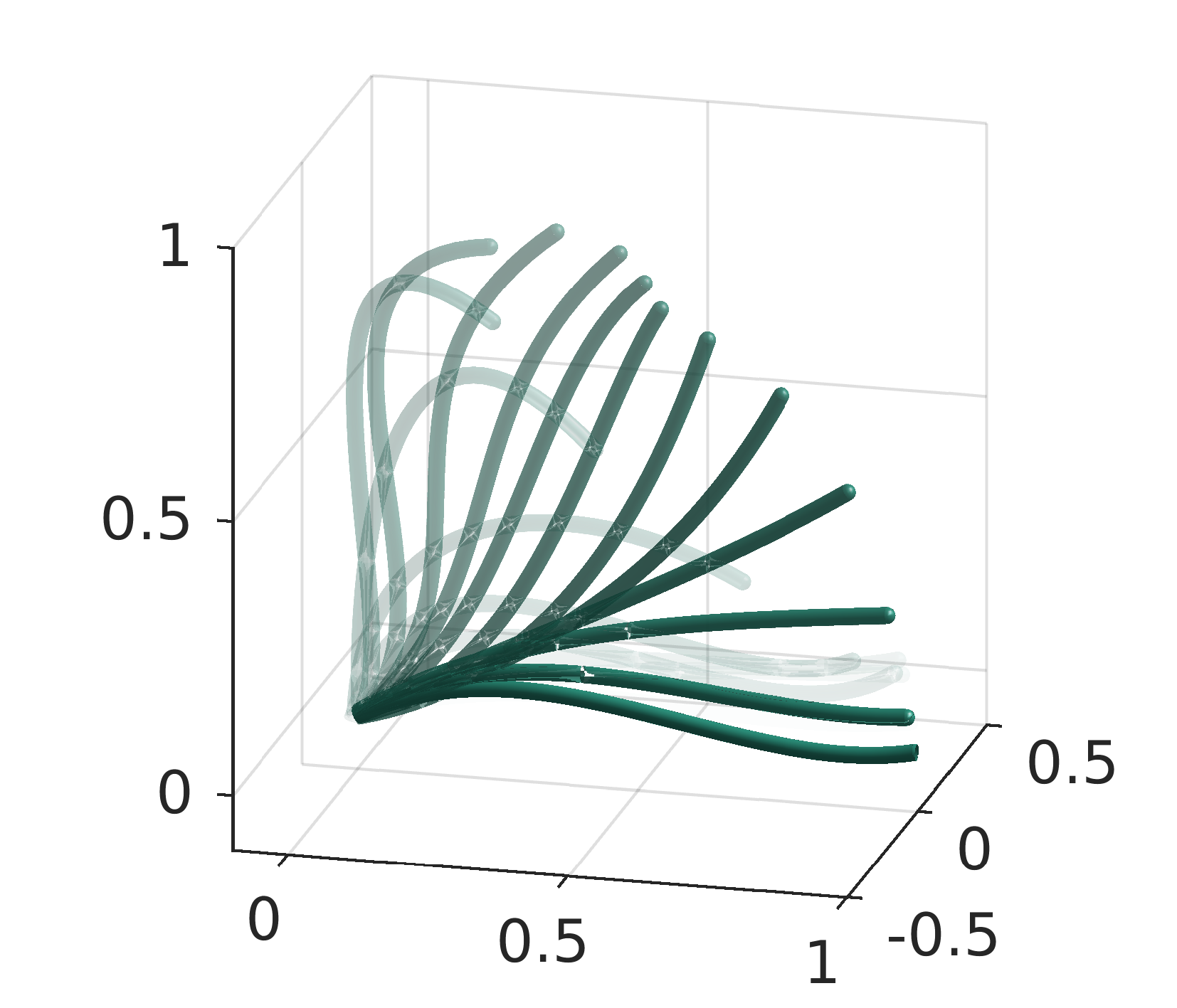}
      &
      \includegraphics[width=7cm]{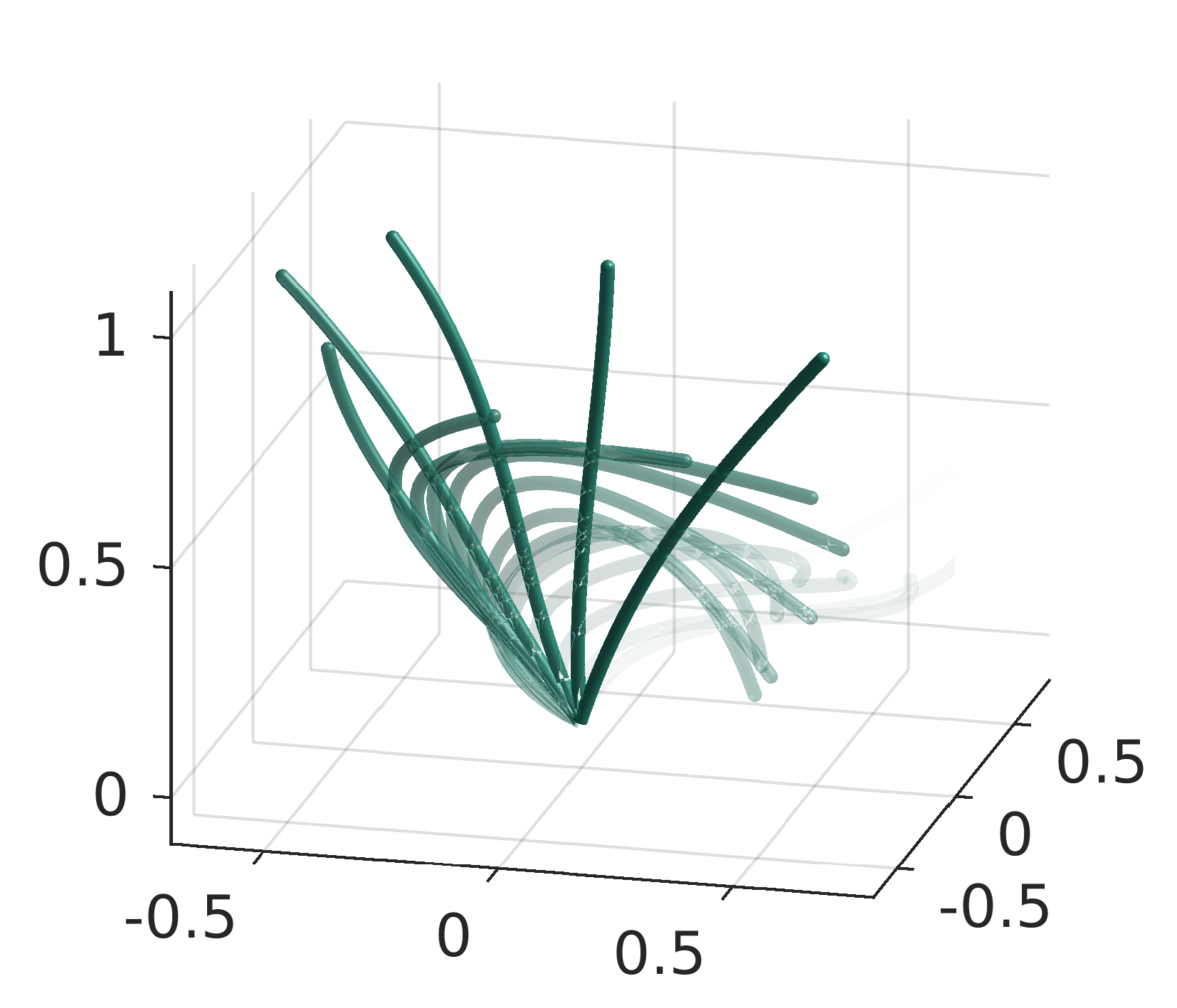}
    \end{tabular}
    \\
    \begin{tabular}{l}
      (c)
      \\
      \includegraphics[width=14cm]{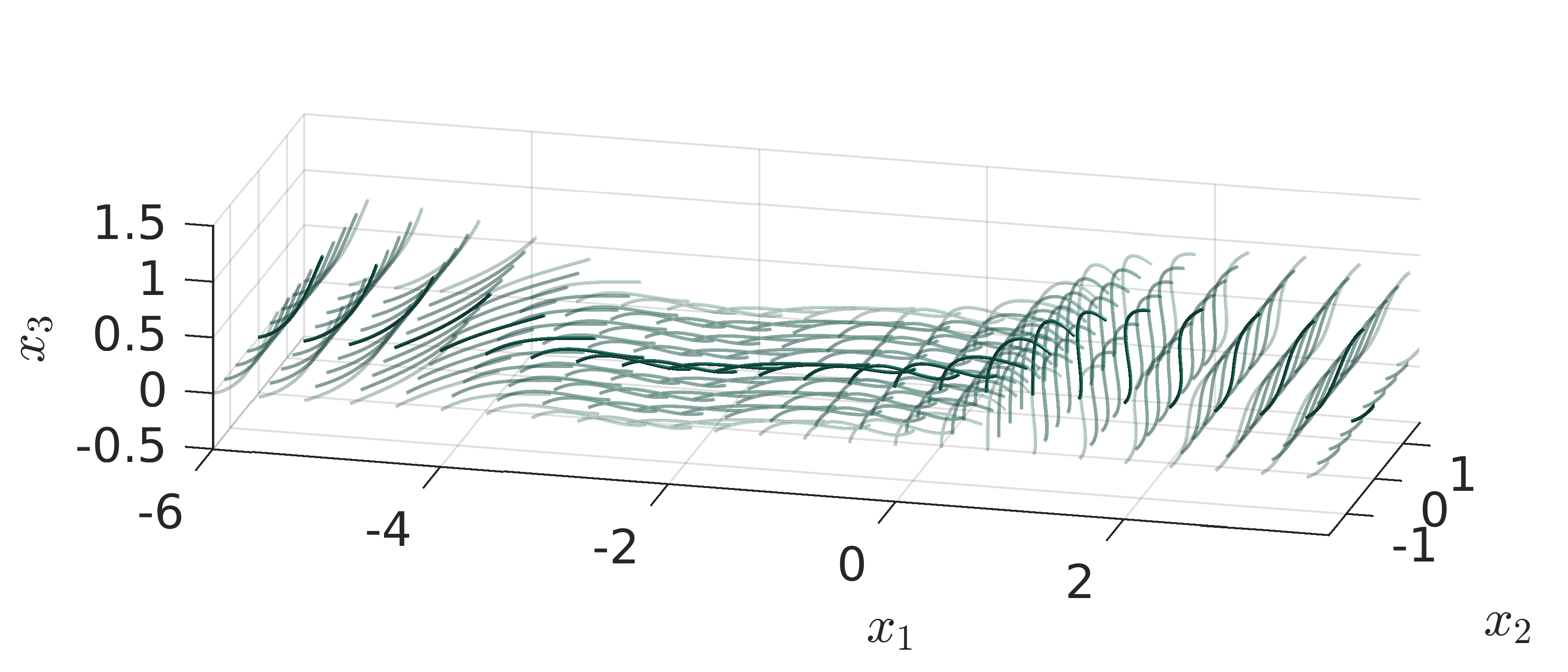}
    \end{tabular}
  \end{tabular}
  \caption{Reconstructed cilia beats from the graphs in ref.~\cite{blake1972model}. (a) \emph{Opalina}, (b) \emph{Paramecium}, (c) the sympletic metachronal wave of an array of \emph{Opalina} cilia.}\label{fig:discreteCiliaBeats}
\end{figure}

The next step was to take into account the fact that cilia form a dense array. Fortunately, the metachronal coordination and periodicity of the ciliary array simplifies the analysis considerably. Mathematically, this coordination can be expressed as,
\begin{equation}
  \begin{aligned}
    \xi_1'(s,t) & = x_1 + \xi_1(s,k x_1 \pm \sigma t), \\
    \xi_2'(s,t) & = x_2 + \xi_2(s,k x_1 \pm \sigma t), \\
    \xi_3'(s,t) & =       \xi_3(s,k x_1 \pm \sigma t), \\
  \end{aligned}
\end{equation}
where \((x_1,x_2,0)\) is the position of the base of the cilium, \(\sigma\) is the radian frequency, positive \(x_1\) is the direction of the effective stroke, and \(+\sigma t\) refers to \emph{antiplectic} metachronism (the wave travels in the opposite direction to the effective stroke), \(-\sigma t\) refers to \emph{symplectic} metachronism (the wave travels in the same direction as the effective stroke). A metachronal wave is depicted in figure~\ref{fig:discreteCiliaBeats}.
Distributing the cilia on a rectangular lattice \(x_1=ma\), \(x_2=nb\) with \(m,n=0, \pm 1, \pm 2, \ldots\), the flow field \(\bm{u}(\bm{x},t)\) can be constructed as a doubly-infinite sum of line integrals,
\begin{equation}
  u_j(\bm{x},t) = \sum_{n=-\infty}^\infty \sum_{m=-\infty}^\infty \int_0^L F_k[\bm{\xi}'] G_{jk}(\bm{x},\bm{\xi}')ds.
\end{equation}
The doubly-infinite sum was then dealt with by using the Poisson summation formula, which replaces the summand with its Fourier transform, and producing exponential convergence. Retaining only the leading order term, and taking both a spatial average in \((x_1,x_2)\) and a temporal average yielded the mean velocity profile as a function of \(x_3\),
\begin{equation}
  \begin{aligned}
    U_\alpha(x_3) & = \frac{1}{\mu a b} \overline{\int_0^L w(s,t) K(x_3,\xi_3)F_\alpha[\bm{\xi}] ds} + O\left(\frac{ab}{L^2}\right) \sigma L \quad (\alpha=1,2), \\
    U_3     (x_3) & = O\left(\frac{ab}{L^2}\right) \sigma L.
  \end{aligned}
\end{equation}
The weight function \(w(s,t)\) depends on the type of metachronism, and the kernel \(K(x_3,\xi_3)\) obtained from spatially integrating \(G_{jk}\) is given by,
\begin{equation}
K(x_3,\xi_3) = \begin{cases} x_3 \quad (x_3 < \xi_3), \\ \xi_3 \quad (x_3 > \xi_3), \end{cases}
\end{equation}
which can be physically interpreted as a shear flow below the singularity location, and a constant streaming flow above.


To calculate flow fields, an essential task was to derive a description of the cilium beat. Cilia beat at or above 10~Hz, are densely-packed, and both their radius and spacing is below the wavelength of visible light. Because of these features, determining the beat pattern was (and still is) challenging -- much more so than determining the beat of sperm flagella, as pursued through high speed cinemicrography by Gray \& Hancock some 15 years earlier. John focused on freehand sketches of M. Sleigh's diagrams of cilia beats of \emph{Opalina}, \emph{Paramecium} and \emph{Pleurobrachia}, derived from high speed cinefilms. Constructing least-squares cubic approximations of each instantaneous waveform, followed by Fourier interpolation over the sequence of waveforms. The procedure was carried out by hand, and in John's words was `not particularly accurate', however they provided the essential characteristics of the beat patterns of different ciliates, as well as providing a methodology which could subsequently be applied to the electron-microscopy data on frozen respiratory cilia \cite{sanderson1981}.

The mathematical form of the model taken was,
\begin{equation}
  \begin{aligned}
    \bm{\xi}(s,\tau) & = \frac{\bm{a}_0}{2} + \sum_{n=1}^N [\bm{a}_n(s)\cos n\tau + \bm{b}_n(s)\sin n\tau], \quad \mbox{with} \\
    \bm{a}_n(s)   & = \sum_{m=1}^M \bm{A}_{mn} s^m, \\
    \bm{b}_n(s)   & = \sum_{m=1}^m \bm{B}_{mn} s^m.
  \end{aligned}\label{eq:fourierbeat}
\end{equation}
Approximations to the coefficients \(\bm{A}_{mn}\), \(\bm{B}_{mn}\) are given in appendix~\ref{app:fourier} and the associated beat patterns depicted in figure~\ref{fig:discreteCiliaBeats}.

The final step in computing flow fields was to model the force. In keeping with the approximate but rational approach, John utilised the resistance coefficient approach of Gray \& Hancock \cite{gray1955}, through which the force on a small segment of a slender body is given by an anisotropic force-velocity relation,
\begin{equation}
  \begin{aligned}
    \delta F_T & = C_T V_T \delta s = C_T (\bm{v}\cdot \bm{t})\delta s, \\
    \delta F_N & = C_N V_N \delta s = C_N (\bm{v}\cdot \bm{n})\delta s,
  \end{aligned}
\end{equation}
where \(\bm{v}\) is the local fluid velocity, the unit vectors \(\bm{t}\) and \(\bm{n}\) are tangent and normal to the flagellum, and \(C_T\), \(C_N\) are resistance coefficients, with \(C_N/C_T \approx 2\), as shown in figure~\ref{fig:disccilia}b. Combined with the spatial and temporal averaging procedure, coupling the velocity field due to a single cilium to the velocity field due to all cilia, and solving numerically, flow profiles could then be computed. The resulting flow profiles showed backward flow close to the surface in the antiplectic species \emph{Paramecium} and \emph{Pleurobrachia}, but not in the symplectic species \emph{Opalina}. Further quantities that were found via this method included the cilium force distributions and rate of working.

The key mechanisms by which protozoan cilia propel fluid would be summarised by John as follows:
\begin{enumerate}
\item The \emph{wall effect} -- cilia are further from the no-slip boundary during the effective stroke than the recovery stroke, and hence propel more fluid.
\item The \emph{Orientation effect} -- cilia move relatively more normal to their axis during their effective stroke, and more tangentially to their axis during their recovery stroke. Because \(C_N>C_T\), more fluid is propelled by the effective motion.
\end{enumerate}

This study, written before John was 25, shows a virtuoso in action, simplifying an intimidating fluid mechanical problem to render it amenable to the limited research computing technology of the late 1960s, and extracting physical insight from the resulting mathematical expressions. As computing technology has developed, some of these simplifications are no longer necessary, however the core ideas of image systems, modelling cilia with singularity solutions, and Fourier representations of the cilia beat are fundamental to current research. In the next section we will see how developments of this first paper could then be applied to human health and disease.



\section{Cilia in physiological flows}
Cilia are found throughout the eukaryotes, in particular being crucial to physiological functions in animals such as respiratory defence, transport of the ovum and embryo, and embryonic left-right symmetry breaking. John's first foray into physiological flow was while he was still at DAMTP, as a postdoctoral fellow \cite{blake1973flow}, formulating a model of flow in ciliated tubules such as the trachea via the methods described in section~\ref{sec:disccilia}. As he had found for protozoan cilia exhibiting antiplectic metachronism, John predicted that reflux could occur close to the tracheal wall. He followed this initial study with a highly-cited paper \emph{On the Movement of Mucus in the Lung} \cite{blake1975movement}, which investigated the effect of cilia, gravity and airflow in transporting mucus, and the effect of the low viscosity peri-ciliary layer which lies below the mucous layer on these mechanisms. In addition to the propulsive mechanisms demonstrated in ciliated protozoa (the wall effect and the orientation effect), respiratory cilia also engage the highly viscous mucous layer only during their effective stroke, the recovery stroke normally taking place entirely in the low viscosity peri-ciliary layer. This selective engagement again breaks forward/backward symmetry, enhancing mucus transport. Other topics in physiological mechanics investigated around this period included flow in the foetal lung \cite{kennewell1976,blake1977foetal} and the role of cilia in ovum transport \cite{blake1983}.

\begin{figure}
  \centering
  \begin{tabular}{c}
    \begin{tabular}{ll}
    (a) & (b)
    \\
  \begin{tikzpicture}
    \draw [->] (-1.5,0) -- (4,0);
    \draw [->] (0,0) -- (0,5.5);
    \draw (0,0) circle (0.05);
    \node at (3.9,-0.3) {\(X_1\)};
    \node at (0.0,-0.3) {\(O\)};
    \node at (0.3, 5.3) {\(X_3\)};
    \draw [<->] (-1.7,0.0) -- (-1.7,3.2);
    \node at (-2.0,1.6) {\(h\)};
    \node at (-0.3,4.5) {\(H\)};
    \node at (3.0,1.6) {\(\ell=1\), \(\mu=\mu_1\)};
    \node at (3.0,4.1) {\(\ell=2\), \(\mu=\mu_2\)};
    \draw [thick] plot [smooth,tension=0.5] coordinates    {(-0.14,0) (-0.09,0.6) (0.21,1.28) (0.56,1.77) (1.38,2.61) (2.34,3.23) (3.22,3.58) (3.30,3.38) (2.46,2.97) (1.62,2.39) (0.84,1.62) (0.45,1.08) (0.16,0.45) (0.14,0)};
    \draw plot [smooth,tension=0.8] coordinates {(-1.5,3.2) (0.0,3.25) (2.00,3.60) (2.83,3.47)};
    \draw plot [smooth,tension=0.8] coordinates {(3.03,3.25) (3.4,3.1) (4.0,3.2)};
    \draw plot [thick,smooth,tension=0.8] coordinates {(-1.5,4.8) (4.0,4.8)};
  \end{tikzpicture}
  &
  \begin{tikzpicture}
    \draw [->] (-1.5,0) -- (4,0);
    \draw [->] (0,0) -- (0,5.5);
    \draw (0,0) circle (0.05);
    \node at (3.9,-0.3) {\(U_1\)};
    \node at (0.0,-0.3) {\(O\)};
    \node at (0.3, 5.3) {\(X_3\)};
    \node at (-0.3,3.2) {\(h\)};
    \node at (-0.3,4.8) {\(H\)};
    \node at (3.0,1.6) {\(\ell=1\)};
    \node at (3.0,4.1) {\(\ell=2\)};
    \draw [dotted] (-1.5,3.2) -- (4,3.2);
    \draw [dotted] (-1.5,4.8) -- (4,4.8);
    \draw plot [smooth,tension=0.3] coordinates {(0,0) (0,1.0) (-0.05,2.4) (3.0,3.0) (3.2,4.0) (3.2,4.8)};
  \end{tikzpicture}
  \end{tabular}
    \\
    \begin{tabular}{l}
    (c)
    \\
  \begin{tikzpicture}
    \node at (0,0) {\includegraphics[trim=1cm 3.4cm 1cm 3.4cm,clip]{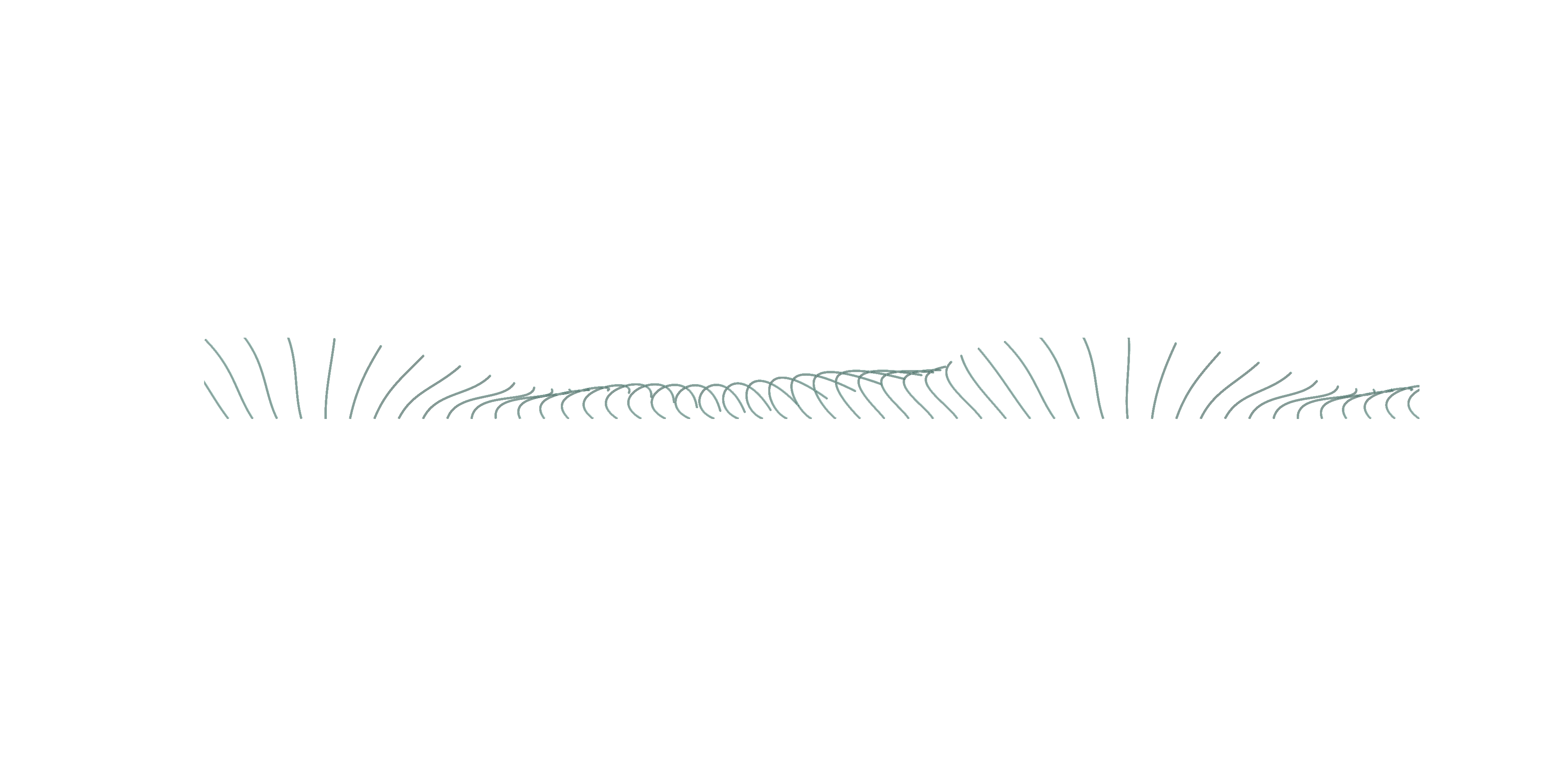}};
    \draw (-6,-0.278) -- (6.5,-0.278);
    \draw [thin,dashed] (-6,0.4) --    (6.5,0.4);
    \draw (-6,0.8) -- (6.5,0.8);
    \draw (6.7,-0.278) -- (6.8,-0.278);
    \draw (6.7,0.4)    -- (6.8,0.4);
    \draw (6.8,-0.278) -- (6.8,0.4);
    \node [anchor=west] at (6.9,0.08) {peri-ciliary layer};
    \node [anchor=west] at (6.9,0.6)  {mucous layer};
    \draw (6.8,0.08) -- (6.9,0.08);
    \draw (6.7,0.8)    -- (6.8,0.8);
    \draw (6.8,0.4)    -- (6.8,0.8);
    \draw (6.8,0.6)    -- (6.9,0.6);
    \node at (0.25,-0.6) {epithelium};
    \node at (0.25, 1.1) {air};
  \end{tikzpicture}
  \end{tabular}
  \end{tabular}
  \caption{Schematic of cilia propelling a two-layer fluid, as found in the respiratory system with a lower watery peri-ciliary layer and upper highly viscous mucous layer \cite{blake1984,fulford1986force,fulford1986}. (a) A single cilium penetrating the mucous layer. (b) Characteristic shape of velocity profiles recovered from the discrete cilia model, with zero or slight backward transport throughout most of the peri-ciliary layer. (c) Depiction of a set of cilia with antiplectic metachronism propelling a two-layer fluid.}\label{fig:twolayer}
\end{figure}
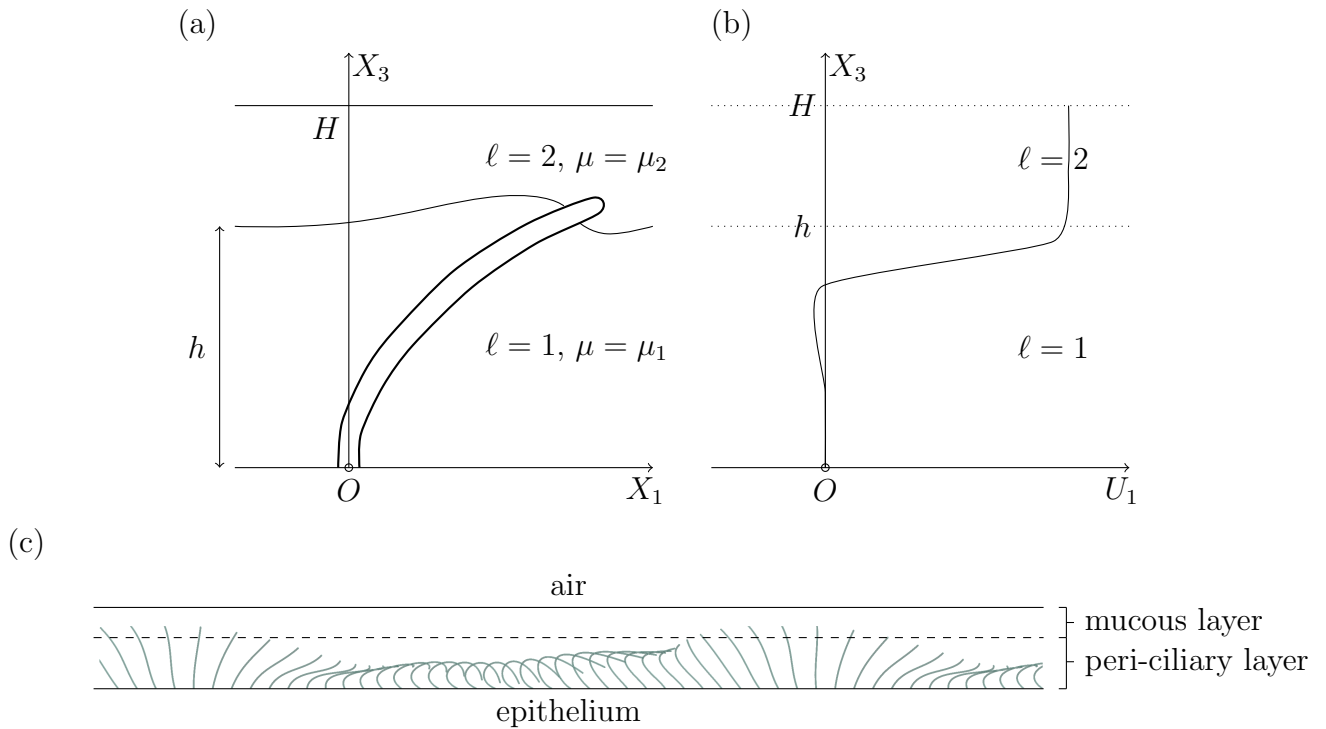

John returned to the subject of respiratory clearance after taking up his Chair at the University of Wollongong (although it should be mentioned that he also collaborated for a period with K.\ Aderogba on the highly relevant topic of the action of a force near a planar two-fluid interface \cite{aderogba1978,aderogba1978add}). With his first PhD student, G.R.\ Fulford, he studied the slender body theory of an object straddling the interface between two viscous fluids (figure~\ref{fig:twolayer}), as occurs during the effective stroke (published in ref.\ \cite{fulford1986force} with a publication date of 1986 but submitted in 1984). Based on this detailed analysis, and retaining the leading order term, a modified theory of the type described in section~\ref{sec:disccilia} could then be derived \cite{blake1984},
\begin{equation}
U_\alpha(x_3,t) = \frac{1}{Nab}\sum_{n=1}^N \int_0^L F_\alpha^{(\ell)}(\bm{\xi}^*) K^{(\ell)}(x_3,\xi_3^*)ds,
\end{equation}
where \(N\) is the number of cilia in one wavelength and the superscript \(\ell\) refers to the peri-cilia layer (\(1\)) or mucous layer (\(2\)), the kernel now taking the form,
\begin{equation}
  \begin{aligned}
    (i) \quad (0<y_3<h) & \\
    K^{(1)}(x_3,y_3) & = \begin{cases} x_3/\mu_1 \quad ; \quad 0\leqslant x_3 < y_3, \\ y_3/\mu_1 \quad ; \quad y_3<x_3\leqslant h, \\ y_3/\mu_1 \quad ; \quad h\leqslant x_3 \leqslant H, \end{cases}
    \\
    (ii) \quad (h<y_3<H) & \\
    K^{(2)}(x_3,y_3) & = \begin{cases} x_3/\mu_1 \quad ; \quad 0\leqslant x_3 \leqslant h, \\ (x_3+(\lambda-1)h)/\mu_2 \quad ; \quad h\leqslant x_3 < y_3, \\ (y_3+(\lambda-1)h)/\mu_2 \quad ; \quad y_3<x_3\leqslant H.\end{cases}
  \end{aligned}
\end{equation}
The theory also revealed behaviour of the interface shape: the interface is depressed immediately in front of the body.

In the follow-up \emph{J.\ Theor.\ Biol.} paper, this theory was applied to a respiratory cilia beat pattern extracted from the electron microscopy data of M.\ Sanderson and M.\ Sleigh \cite{sanderson1981}, leading to a prediction of optimal cilia penetration depth (10--20\%) of the cilium length, and the conclusion that cilia penetration was necessary for effective transport only in the case of low ciliary activity. Additionally, it was confirmed that mean transport of the peri-ciliary fluid was likely to be minimal by comparison with the mucous layer. These studies culminated in a highly-cited review \emph{State of Art} review with Sleigh and N.\ Liron \cite{sleigh1988}.

Over a decade later, experiments on human tracheo-bronchial cultures conducted at the University of Carolina Chapel Hill would appear to refute the finding of minimal peri-ciliary fluid transport, in part of a study on the competing roles of fluid depletion and tonicity on the airway disease that occurs in cystic fibrosis \cite{matsui1998}. By photo-uncaging fluorescent molecules dispersed through the surface liquid and visualising with confocal microscopy, the Chapel Hill team appeared to have observed `cotransport' of both layers, contradicting the velocity profile prediction of the type shown in figure~\ref{fig:twolayer}b. These experiments would re-kindle John's interest in the respiratory system, initially through a conference paper with Eamonn Gaffney \cite{blake2001modeling}, followed by my PhD research with John and Eamonn which commenced at the beginning of 2002. My work with John and Eamonn focused on the role of viscoelastic effects and pressure gradients induced by the presence of the mucous layer, formulated by developing a spatially-continuous `traction layer' model \cite{smith2007viscoelastic}, building on the active porous medium model John developed while at CSIRO in the late 1970s \cite{blake1977}. The combination of large oscillations in the velocity field induced by the cilia beat with the effect of shear-enhanced diffusion across the thin airway surface liquid layer resulted in predictions of fluorescent molecule transport that were compatible with the Chapel Hill experiments, even with minimal time-averaged peri-ciliary liquid velocity \cite{smith2007model}. The revisited model also enabled effects such as airway surface liquid depth and viscosity ratio to be investigated further -- an initially surprising prediction of the model was that reducing the viscosity of the mucus may actually increase transport, provided that the mucus-peri-cilia interface is maintained -- however this prediction was supported by the fact that the heavily-diluted airway surface liquid transport occurring in the condition pseudohypoaldosteronism is transported more rapidly than normal. These findings are summarised in more detail in the review \cite{smith2008modelling}.

John also encouraged me to revisit discrete cilia modelling, an area in which it was now possible to make further progress because of the increased computational power that had become available over the previous two decades. We were able to model flow around cilia below an oscillating mucous layer \cite{smith2007discrete}, combining John's earlier modelling ideas with methodological developments of three earlier collaborators of John's, N.\ Liron, A.\ Chwang and T.Y.\ Wu \cite{liron1978,chwang1975}, and M.\ Staben et al.'s repurposing of the Blake image system \cite{staben2003}. We later applied our work on discrete cilia modelling to the embryonic node, as will be discussed in section~\ref{sec:nodal}.

\section{Flagellates, blinking stokeslets and chaotic advection}

Another important role of cilia in nature is feeding. Examples which John published on include filter feeding of bivalve mussels \cite{blake1995} and sessile ciliated aquatic microorganisms such as \emph{Vorticella} and \emph{Stentor}. The latter example led him, working with Steve Otto, a colleague at the University of Birmingham, to the idea of the \emph{blinking stokeslet}.

The earlier study of J.J.L.\ Higdon \cite{higdon1979} (also a student of Lighthill) predicted that the feeding current generated by a flagellum produced only a toroidal eddy, which would not be optimal for bringing a continuous supply of new nutrients to the feeder. Reasoning that sessile microorganisms must be able to increase the volume which they can sample, John postulated that these microorganisms may alter the length of their stalk in order to produce a chaotic flow field. To demonstrate the idea, Blake \& Otto \cite{blake1996,blake1998} constructed a conceptually simple, but dynamically rich blinking stokeslet model, an adaptation of the earlier \emph{blinking vortex} of Aref \cite{aref1984}.

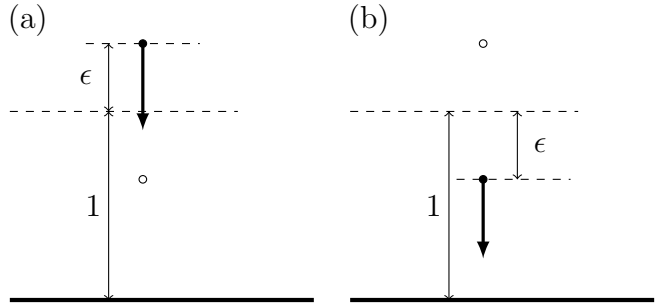
\begin{figure}
  \centering
  \begin{tabular}{ll}
    (a)
    &
    (b)
    \\
    \begin{tikzpicture}
      \draw [ultra thick] (0,0) -- (4,0);
      \draw [dashed] (0,2.5) -- (3,2.5);
      \draw [dashed] (1,3.4) -- (2.5,3.4);
      \filldraw (1.75,3.4) circle (0.05);
      \draw [-latex,very thick] (1.75,3.4) -- (1.75,2.25);
      \draw [<->] (1.3,2.5) -- (1.3,3.4);
      \node at (1.0,2.95) {\(\epsilon\)};
      \draw [<->] (1.3,0.0) -- (1.3,2.5);
      \node at (1.1,1.25) {\(1\)};
      \draw (1.75,1.6) circle (0.05);
    \end{tikzpicture}
    &
    \begin{tikzpicture}
      \draw [ultra thick] (0,0) -- (4,0);
      \draw [dashed] (0,2.5) -- (3,2.5);
      \draw [dashed] (1.4,1.6) -- (2.9,1.6);
      \draw (1.75,3.4) circle (0.05);
      \draw [-latex,very thick] (1.75,1.6) -- (1.75,0.55);
      \draw [<->] (2.2,2.5) -- (2.2,1.6);
      \node at (2.5,2.05) {\(\epsilon \)};
      \draw [<->] (1.3,0.0) -- (1.3,2.5);
      \node at (1.1,1.25) {\(1\)};
      \filldraw (1.75,1.6) circle (0.05);
    \end{tikzpicture}
  \end{tabular}
  \caption{Schematic of the \emph{blinking stokeslet} model of microorganism feeding. The flow is produced by a switching protocol between point forces acting at \(\bm{\xi}=[0,1+\epsilon]\) and \(\bm{\xi}=[0,1-\epsilon]\), both pointing towards the boundary.} \label{fig:blinkingstokeslet}
\end{figure}

The blinking stokeslet is depicted in figure~\ref{fig:blinkingstokeslet}. Briefly, the effect of the microorganism on the fluid is assumed to switch between two states -- concentrated forcing of fluid towards the boundary at height \(h=1+\epsilon\), and concentrated forcing of fluid towards the boundary at height \(h=1-\epsilon\). In 2D, the flow due to a stokeslet pointing towards a plane boundary is given by,
\begin{equation}
  u_x =\frac{\partial \Psi}{\partial y}, \quad u_y = -\frac{\partial \Psi}{\partial x},
\end{equation}
where the streamfunction \(\Psi\) has the form,
\begin{equation}
  \Psi(x,z) = \frac{F}{8\pi\mu}x\left[\frac{1}{2}\ln\left\{\frac{x^2+(y+h)^2}{x^2+(y-h)^2}-\frac{2hy}{(x^2+(y-h)^2}\right\}\right].
\end{equation}
The streamfunction for the blinking stokeslet is then,
\begin{equation}
\Psi(x,y;t) = \Omega_+(t) \Psi_+(x,y)+\Omega_-(t) \Psi_-(x,y),
\end{equation}
with the switching protocol defined by,
\begin{equation}
\Omega_{\pm} = 2 \Omega_{\pm}^0 \sin\left(\frac{t\pi}{\tau}\right)\left\{\sin\left(\frac{t\pi}{\tau}\right)\pm\left|\sin\left(\frac{t\pi}{\tau}\right)\right|\right\}.
\end{equation}
With fixed force location, particles would move along the closed streamlines shown in figure~\ref{fig:images}d, however the temporally-continuous periodic switching of the force location caused particles to move between streamlines, greatly increasing mixing and hence sampling space. The existence of chaotic advection was explored via Poincar\'e sections, as shown in figure~\ref{fig:poincare}. A key finding was that, as the period of the switching oscillation \(\tau\) increased, the flow became more chaotic.

In a follow-up paper with postdoctoral researcher A.N.\ Yannacopoulos, the analysis of chaotic advection was further developed through the introduction of white noise, to model molecular diffusion, and with additional mathematical tools, including delta-function temporal switching, enabling the construction of an implicit map, and the calculation of finite-time Lyapunov exponents \cite{otto2001}, revealing remarkably beautiful and complex dynamics from such a conceptually simple model. Further applications of this idea included the feeding of the choanoflagellate \emph{S.\ Amphoridium} \cite{orme2001}; a review following a 2011 meeting in Leiden which John participated in along with H.\ Aref and many other leading figures in the field has recently been published as ref.\ \cite{aref2017}.

\begin{figure}
  \caption{Poincar\'e sections produced by a blinking stokeslet, for two different switching periods \(\tau\). (a) \(\tau=0.1\), (b) \(\tau=0.5\). From ref.~\cite{blake1998} \emph{figure excluded from arxiv version}.}\label{fig:poincare}
\end{figure}

\begin{figure}
  \centering
  \includegraphics{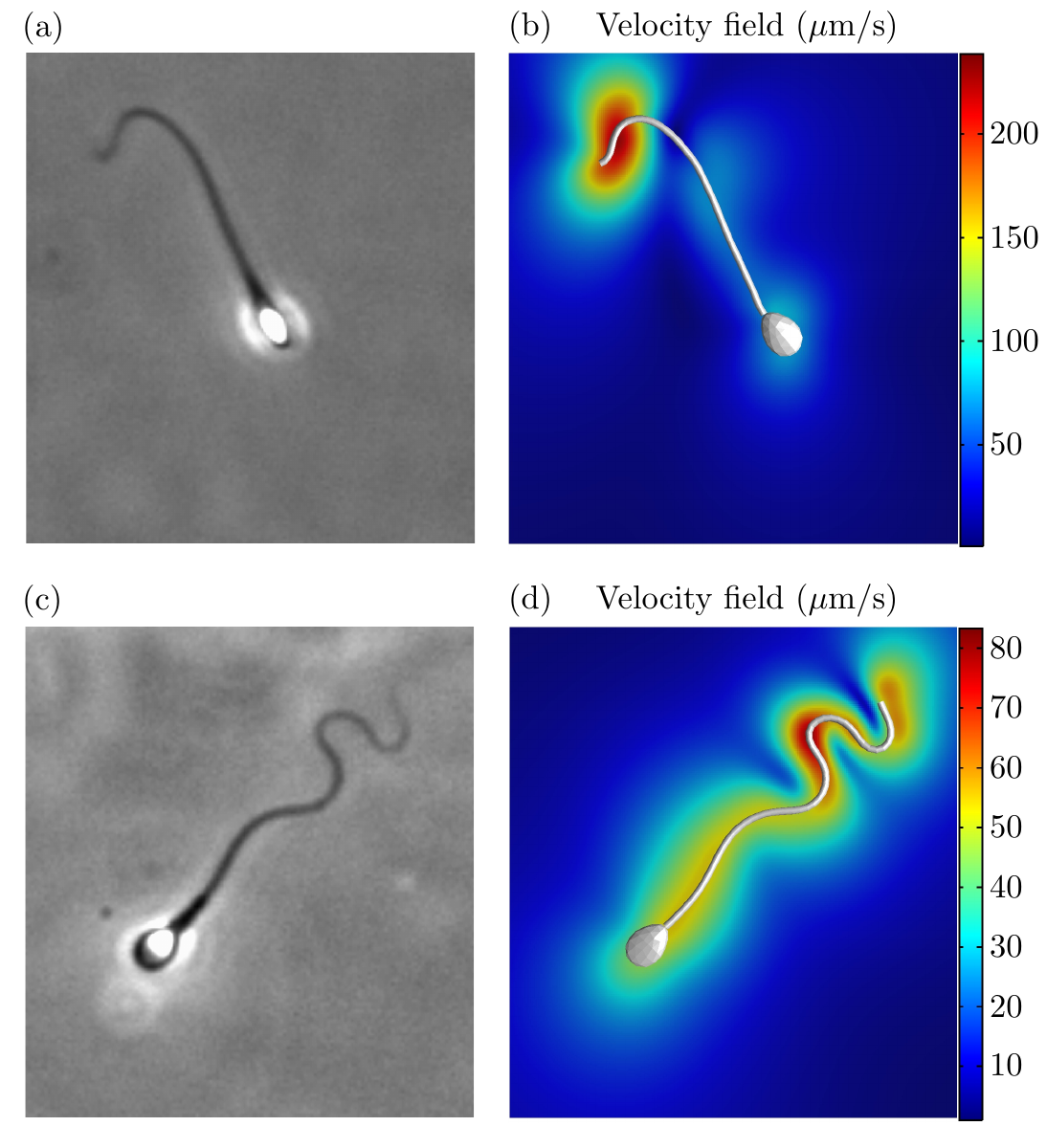}
  \caption{The flow velocity magnitude around a swimming human sperm, via similar methods to ref.\ \cite{gaffney2011}. (a,b) A cell in low viscosity fluid (e.g.\ laboratory saline), (c,d) a cell which has penetrated fluid with viscosity similar to midcycle mucus (\(0.14\)~Pa.s). (a,c) high-speed imaging; (b,d) flow field simulation using waveforms captured from experiment, calculated from slender body theory and the boundary element method. Through its altered beat pattern, the cell in high viscosity fluid produces a much smaller magnitude flow, however it progresses through fluid at the same progressive velocity of approximately \(50\)~\(\mu\)m/s. The absence of a `wake' emphasises the very low Reynolds nature of the flow regime in both cases.}\label{fig:spermfig}
\end{figure}

Another type of \emph{flagellate} is the sperm cell of higher animals, including humans. John brought together a group of younger researchers involving Birmingham Women's NHS Foundation Trust Assisted Conception Unit, the School of Medicine, and Mathematics at the University of Birmingham, including myself. This team has grown over the last decade to include the Universities of Oxford, York, Warwick and Kyoto and has led to a number of publications in journals ranging from applied mathematics and physics to reproductive medicine. During the period of John's leadership up to around 2011, our findings included a fluid dynamic interpretation of the surface accumulation effect in sperm \cite{smith2009surface,smith2009human}, exploration of the flow field (figure~\ref{fig:spermfig}) and internal moment generation in motile cells in physiological viscosity fluid \cite{gaffney2011}. John certainly had an excellent instinct for the `right problem' to focus on -- I will conclude with perhaps his favourite example.

\section{Embryonic nodal cilia}\label{sec:nodal}

In July 2002, relatively early in my PhD studies and shortly after being inducted into the modelling techniques described in section~\ref{sec:disccilia}, John showed me a Nature Letter \cite{nonaka2002} by the Hamada group in Osaka, Japan. This paper addressed the role of cilia-driven flow in the early stages of embryonic left-right symmetry breaking. The left-right axis in mammalian (including human) embryo development is the last to appear, and in the vast majority of individuals breaks the same way -- the heart is to the left of the mid-line, the liver to the right. The role of cilia in this process had been long suspected, following Afzelius' discovery of defects in cilium structure in individuals with Kartagener's syndrome (a triad of respiratory disease, male infertility, and transposed left-right \emph{situs}) in the early 1970s \cite{berdon2004situs,berdon2004more}, but it was not until around 20 years later that cilia were discovered in the embryonic node of the mouse at approximately 8 days post-fertilisation \cite{sulik1994}, along with the existence of the \emph{nodal flow} \cite{nonaka1998}. The process is shown schematically in figure~\ref{fig:nodal}: a cilia-driven leftward flow is produced lower in the layer (this is conventionally drawn the wrong way round), with a slower rightward flow induced higher up in the layer, induced by the presence of the overlying Reichert's membrane, and associated mass conservation.

The 2002 paper of Nonaka et al.\ \cite{nonaka2002} described remarkable experiments in which the flow was reversed artificially in \emph{ex vivo} mouse embryos, resulting in reversal of situs in normal embryos, and correction of situs in mutant embryos with immotile cilia. Our initial modelling efforts did not however result in an explanation for the nodal flow -- nodal cilia were reported to `whirl' clockwise when viewed from above, but how this flow was converted into directional transport was unclear. Two years later, J.\ Cartwright, I.\ Tuval and O.\ Piro demonstrated conceptually that a whirling cilium which was tilted towards the already-established posterior axis would break symmetry, and they calculated the flow field associated with an array of tilted rotlets. The tilt was confirmed experimentally soon afterwards \cite{okada2005}, and modelled experimentally \cite{nonaka2005}. The latter paper re-ignited John's interest in the area, and so as I was finishing my PhD studies, he encouraged me to revisit the problem, making use of the image systems he had derived himself as a PhD student.

Modelling individual and small groups of tilted, whirling cilia -- the cilia in the node are sparser than in the respiratory epithelium -- we simulated the transport of morphogen-containing parcels. As may be expected, close to the cilia the flow was highly vortical, with particles making multiple orbits before `escaping' towards the left; above the cilia tips, particles underwent a `loopy drift' to the left. The latter behaviour is not obvious from a `pure rotlet' model, however it can be interpreted as resulting from the image system induced by the no-slip boundary. The stokeslet associated with a parallel-oriented point force induces an \(O(hr^{-2})\) stresslet far-field, whereas the perpendicular-oriented point force decays more rapidly, so can be neglected to a first approximation. The rotational beat cycle can therefore be broken up into a rightward motion, and a leftward motion, in the latter case with \(h\) ranging over larger values: the net effect is a leftward drift \cite{smith2007discrete}: an example of the wall effect John had found in his work on ciliates as a PhD student. A similar `time averaged' interpretation can be deduced from the image system associated with a point torque oriented parallel to a plane boundary (\cite{blake1974fundamental,smith2007discrete}). These qualitative features were soon afterwards observed experimentally in the zebrafish embryo through imaging particles released by laser ablation \cite{supatto2008}.

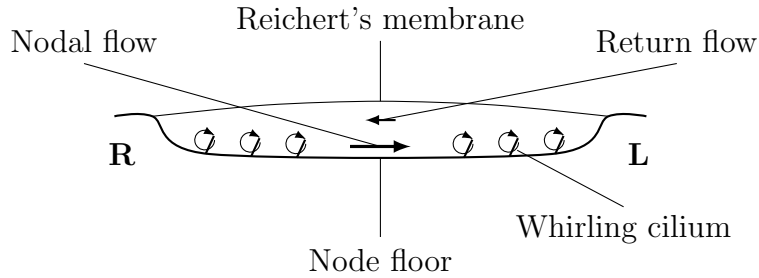
\begin{figure}
  \centering
  \begin{tikzpicture}
    \draw [thick] plot [smooth] coordinates {(0,0.5) (0.5,0.5) (1.25,0.0) (5.75,0.0) (6.5,0.5) (7.0,0.5)};
    \draw         plot [smooth] coordinates {(0.5,0.5) (2.0,0.65) (3.5,0.7) (5.0,0.65) (6.5,0.5)};
    \draw (3.5,0.7) -- (3.5,1.5);
    \node at (3.5,1.8) {Reichert's membrane};
    \draw (3.5,-0.05) -- (3.5,-1.1);
    \node at (3.5,-1.4) {Node floor};
    \draw [-latex,very thick] (3.1,0.1) -- (3.9,0.1);
    \draw (3.5,0.1) -- (-0.4,1.2);
    \node at (-0.4,1.5) {Nodal flow};
    \draw [-latex,thick] (3.7,0.45) -- (3.3,0.45);
    \draw (3.5,0.45) -- (7.4,1.2);
    \node at (7.4,1.5) {Return flow};
    \node at (0.1,0.0) {\textbf{R}};
    \node at (6.9,0.0) {\textbf{L}};
    \draw [thick] (1.2,0.0) -- (1.31,0.24);
    \draw (1.3,0.25) arc (30:360:0.13);
    \draw [-latex] (1.27,0.275) -- (1.32,0.25);
    \draw [thick] (1.8,-0.03) -- (1.91,0.21);
    \draw (1.9,0.22) arc (30:360:0.13);
    \draw [-latex] (1.87,0.245) -- (1.92,0.22);
    \draw [thick] (2.4,-0.05) -- (2.51,0.19);
    \draw (2.5,0.20) arc (30:360:0.13);
    \draw [-latex] (2.47,0.225) -- (2.52,0.20);
    \draw [thick] (5.8,0.0) -- (5.91,0.24);
    \draw (5.9,0.25) arc (30:360:0.13);
    \draw [-latex] (5.87,0.275) -- (5.92,0.25);
    \draw [thick] (5.2,-0.03) -- (5.31,0.21);
    \draw (5.3,0.22) arc (30:360:0.13);
    \draw [-latex] (5.27,0.245) -- (5.32,0.22);
    \draw [thick] (4.6,-0.05) -- (4.71,0.19);
    \draw (4.7,0.20) arc (30:360:0.13);
    \draw [-latex] (4.67,0.225) -- (4.72,0.20);
    \draw (5.3,0.05) -- (6.7,-0.7);
    \node at (6.7,-1.0) {Whirling cilium};
  \end{tikzpicture}
  \caption{Schematic of the embryonic node of the mouse -- note the convention of depicting the left of the embryo (\textbf{L}) on the right hand side of the figure.}\label{fig:nodal}
\end{figure}

We later (see appendix~\ref{app:qformula} for a more personal account) explored the volume flow rate produced by a single tilted rotating cilium in the vicinity of a plane boundary, to attempt to deduce to what extent the cilium beat pattern is optimised to produce maximum flow. John derived a very simple formula for this, which belies a fairly involved derivation (we later arrived at a somewhat easier derivation through some geometrical intuition, see \cite{smith2011}). For a cilium of length \(L\), projecting from a plane boundary at \(x_3=0\),  performing a conical rotation (clockwise viewed from above) with semicone angle \(\psi\), tilted by angle \(\theta\) in the posterior direction, and radian frequency \(\omega\), in a fluid with dynamic viscosity \(\mu\), the time-averaged volume flow rate in the \(x_1\) direction \(Q\) is given by,
\begin{equation}
Q = \frac{C_N \omega L^3}{6\pi\mu} \sin^2 \psi \sin \theta, \label{eq:qformula}
\end{equation}
where \(C_N\) is the resistance coefficient associated with normal motion of the cilium. The geometric set up is shown in figure~\ref{fig:nodal}a, where \(x_1)\) is the right-left axis, \(x_2\) is dorsal-ventral and \(x_3\) is posterior-anterior (in all cases, negative-to-positive). The angular dependence is shown in figure~\ref{fig:nodal}b. Noting that \(\psi+\theta\leqslant \pi/2\) and that as a non-decreasing function \(Q(\psi,\theta)\) has its maximum value on the boundary \(\psi+\theta=\pi/2\), we then seek the maximum of \(\sin^2\psi \sin(\pi/2-\psi)\), which is easily found to be \(\psi=\mathrm{arctan} \sqrt{2}\approx 54.7^\circ\), with \(\theta\approx 35.3^\circ\). I (and colleagues) were initially surprised that the optimal angle was not simply \(\psi=\theta=45^\circ\) -- although this was clear from the mechanical experiments of Nonaka et al.\ \cite{nonaka2005}. The formula compared well to more precise simulation based on a computational implementation of slender body theory, and the optimal angle prediction of \(\theta\approx 35^\circ\) was consistent with the (fairly wide range) of values reported in experiment -- \(40^\circ\) by Okada et al.\ \cite{okada2005} to \(27^\circ\) by Nonaka et al.\ \cite{nonaka2005}.

Equation~\eqref{eq:qformula} has since been extended to take account of higher order multipoles \cite{vilfan2012}, modified to produce a parameterised point-torque reprepresentation for whole-organ modelling \cite{montenegro2016}, and generalised to model helical waveforms in long-cilia mutants \cite{pintado2017}. John told me that his `\(Q\)-formula', derived from coordinate geometry, calculus, resistive force theory, and the Blake image system, to produce a physical insight about the origins of symmetry in vertebrate embryos, was one of his proudest career achievements. Its elegance, utility and generalisability exemplify his approach to science.

\begin{figure}
  \centering
  \begin{tabular}{ll}
    (a)
    &
    (b)
    \\
    \begin{tikzpicture}
      \node at (0,0) {\includegraphics{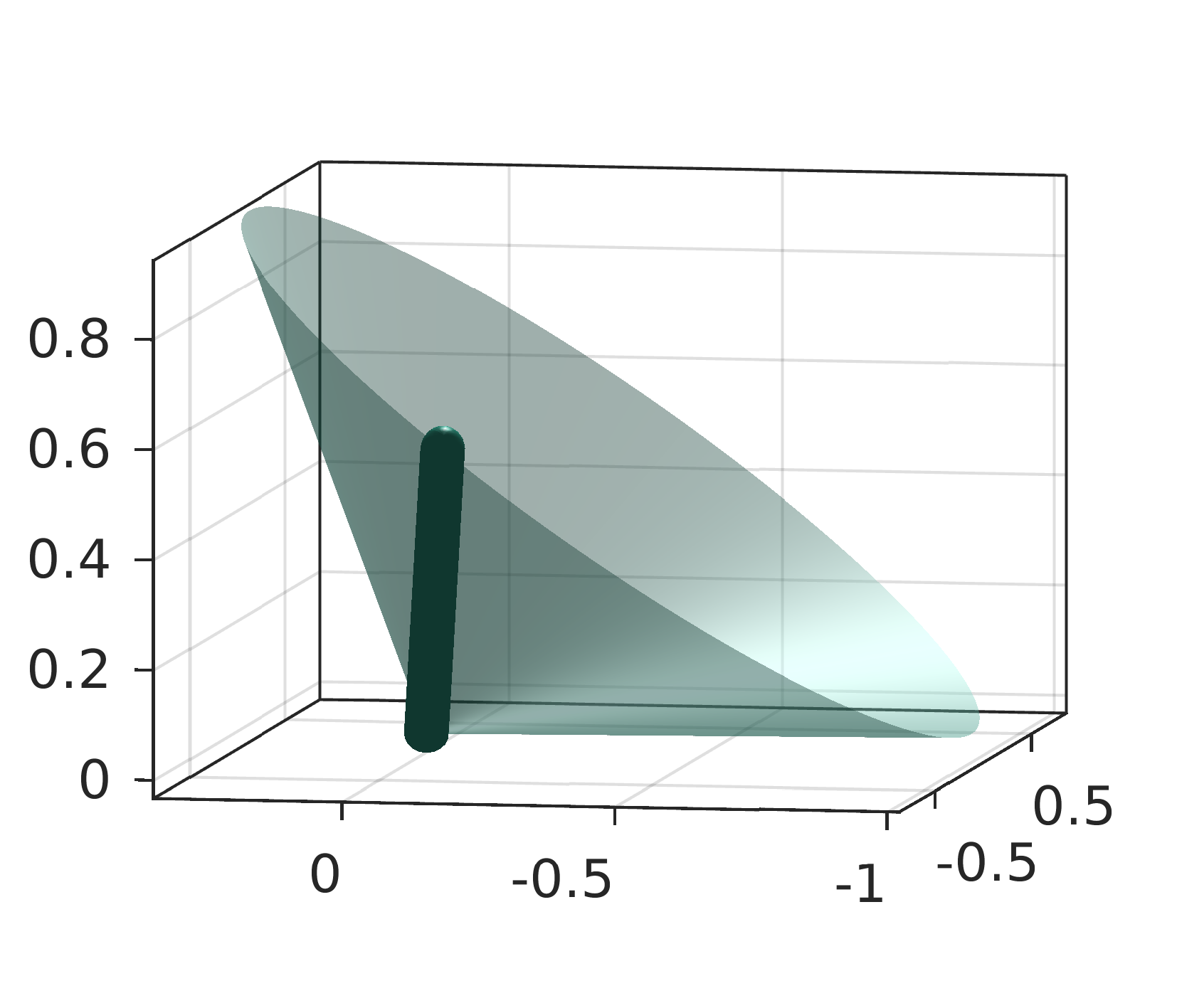}};
      \draw [dashed] (-0.94,-1.37) -- (-0.94,1.9);
      \draw [dashed] (-0.94,-1.37) -- (1.28,1.9);
      \draw (-0.94,-0.37) arc (90:55:1);
      \node at (-0.68,-0.65) {\(\theta\)};
      \draw (-0.49,-0.72) arc (55:0:0.8);
      \node at (-0.49,-1.1) {\(\psi\)};
    \end{tikzpicture}
    &
    \begin{tikzpicture}
      \node at (0,0) {\includegraphics{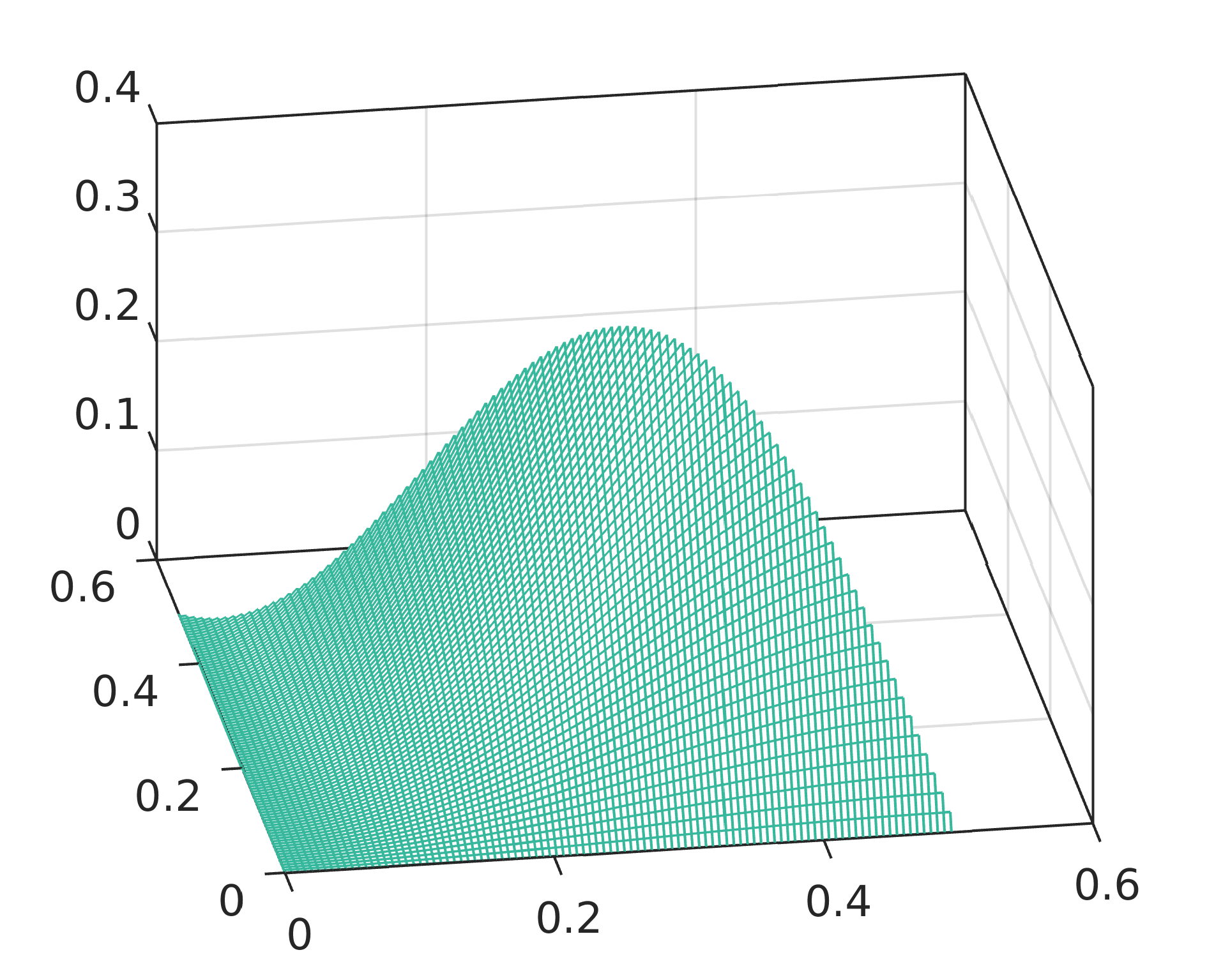}};
      \node at (0.8,-3.2) {\(\psi/\pi\)};
      \node at (-3.8,-1.9) {\(\theta/\pi\)};
    \end{tikzpicture}
  \end{tabular}
  \caption{Geometric model of a whirling nodal cilium and prediction of the dependence of flow rat from the geometric parameters. (a) A model of a nodal cilium as a rigid whirling rod, tracing out a conical envelope with semi-cone angle \(\psi\) and tilt angle \(\theta\) towards the posterior. (b) Plot of the function \(\sin^2\psi\sin \theta\) for the angles \(0\leqslant \psi \leqslant \pi/2\) and \(0\leqslant \theta + \psi \leqslant \pi/2\). The optimum occurs for \(\psi+\theta=\pi/2\) and \(\psi=\mathrm{arctan}\sqrt{2}\approx 54.7^\circ\).}\label{fig:nodal}
\end{figure}
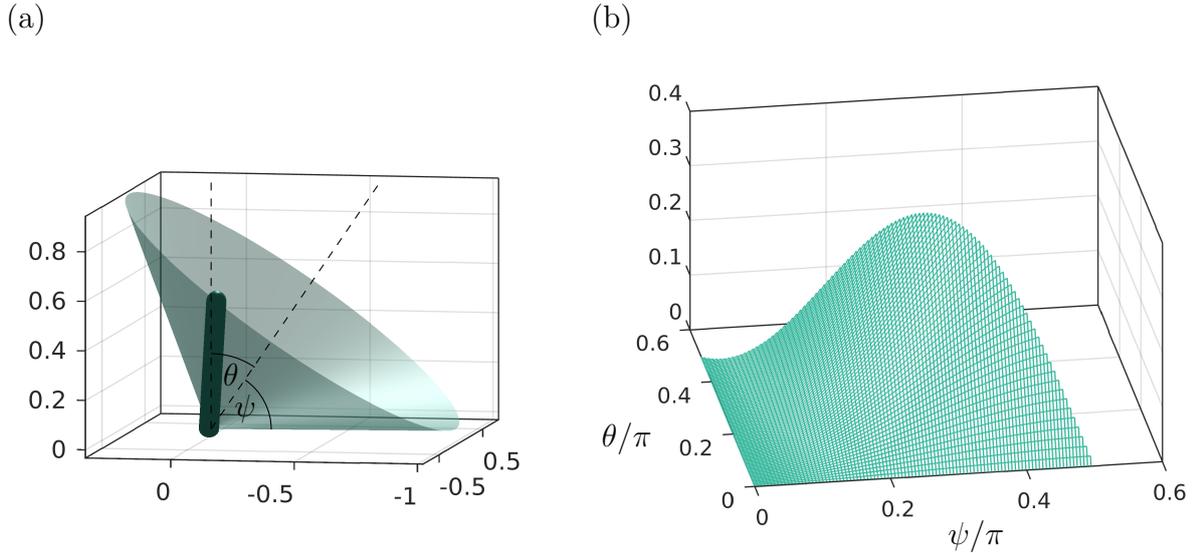

\section{Conclusions and personal reflection}

This article has highlighted some of the many areas of biological fluid mechanics to which John Blake made major contributions, in addition to a personal perspective to give an indication of the inspiration he provided to his students. John supervised about 22 PhD students and 17 postdoctoral researchers -- I was fortunate to work with him in both capacities, over a period (2002--2013) during which biological fluid mechanics was making a resurgence, with many researchers internationally building on John's achievements. His enthusiasm, wealth of knowledge, insight and resources provided a wonderful intellectual environment for an aspiring scientist.

If I may attempt to characterise John's scientific ethos, it would be that he developed elegant mathematical approaches to reveal the fundamental physical effects in complicated living systems. He combined the ability to distill the essential physics of a system mathematically, with an instinct for the most fertile ground for applied mathematicians to collaborate with experimentalists. John also insisted on the importance of disseminating findings to all of the communities that needed to hear about them.

In summary, John Blake provided mathematical theory to understand and make predictions about the microscale world of fluid mechanics around cells, enabling us to use mathematics as an \emph{improved microscope}, capable of seeing the mechanical interactions and principles of movement and transport as well as shape and form. His legacy is not simply the techniques he developed but also the collaborations he fostered, the ethos he championed, and the scientific careers he nurtured.

\begin{appendices}

\section{Fourier coefficients of the reconstructed cilia beat patterns}\label{app:fourier}
The coefficients \(\bm{A}_{mn}\), \(\bm{B}_{mn}\) of equation~\eqref{eq:fourierbeat} were not published in ref.\ \cite{blake1972model} (although coefficients were reported for subsequent studies on respiratory tract cilia). Retracing John's steps by performing the Fourier least-squares fit with Matlab leads to the coefficients given in table~\ref{tab:fourier}. These coefficients are somewhat approximate, however I recall John's amusement at younger researchers taking to excessive precision coefficients derived with, in his words, a `sheet of acetate and a piece of string'!

\begin{table}[h]
  \centering
  \begin{tabular}{c}
    \begin{tabular}{c}
    (a) \emph{Opalina}
    \\[1em]
      \begin{tabular}{cccccc}
      \multicolumn{6}{c}{\(\bm{A}_{mn}\)} \\[0.5em]
      \(n=\) & \(0\) & \(1\) & \(2\) & \(3\) & \(4\)\\[0.5em]
      \(m =\) 1 & \(  0.725\) & \(  0.097\) & \(  0.187\) & \( -0.052\) & \( -0.005\) \\
      & \( -0.634\) & \( -0.009\) & \(  0.112\) & \(  0.229\) & \( -0.095\) \\[0.5em]
      \(m =\) 2 & \( -0.223\) & \(  0.849\) & \( -0.615\) & \( -0.043\) & \(  0.115\) \\
      & \(  0.150\) & \(  1.206\) & \( -0.191\) & \( -0.790\) & \(  0.379\) \\[0.5em]
      \(m =\) 3 & \(  0.204\) & \( -0.593\) & \(  0.393\) & \(  0.029\) & \( -0.078\) \\
      & \(  0.026\) & \( -0.769\) & \(  0.053\) & \(  0.553\) & \( -0.285\) \\[1em]
      \multicolumn{6}{c}{\(\bm{B}_{mn}\)} \\[0.5em]
      \(n=\)  & &  \(1\) & \(2\) & \(3\) & \(4\)\\[0.5em]
      \(m=1\) & & \( -0.336\) & \(  0.145\) & \(  0.089\) & \( -0.094\) \\
      & & \( -0.298\) & \( -0.012\) & \(  0.137\) & \(  0.091\) \\[0.5em]
      \(m=2\) & & \(  0.264\) & \(  0.186\) & \( -0.387\) & \(  0.121\) \\
      & & \(  0.387\) & \(  0.389\) & \( -0.432\) & \( -0.497\) \\[0.5em]
      \(m=3\) & & \(  0.007\) & \( -0.254\) & \(  0.251\) & \( -0.029\) \\
      & & \( -0.002\) & \( -0.398\) & \(  0.206\) & \(  0.398\)\\[2em]
      \end{tabular}
    \\
    (b) \emph{Paramecium}
    \\[1em]
      \begin{tabular}{cccccc}
      \multicolumn{5}{c}{\(\bm{A}_{mn}\)} \\[0.5em]
      \(n=\) & \(0\) & \(1\) & \(2\) & \(3\)\\[0.5em]
      \(m =\) 1 & \( -0.583\) & \(  0.436\) & \( -0.456\) & \( -0.021\) \\
      & \( -0.800\) & \(  0.084\) & \( -0.090\) & \(  0.025\) \\[0.5em]
      \(m =\) 2 & \(  1.150\) & \( -2.492\) & \(  1.012\) & \(  0.279\) \\
      & \( -0.164\) & \( -0.286\) & \(  0.294\) & \(  0.192\) \\[0.5em]
      \(m =\) 3 & \( -0.276\) & \(  1.645\) & \( -0.928\) & \( -0.358\) \\
      & \(  0.453\) & \( -0.132\) & \( -0.240\) & \( -0.141\) \\[1em]
      \multicolumn{4}{c}{\(\bm{B}_{mn}\)} \\[0.5em]
      \(n=\)  & & \(1\) & \(2\) & \(3\)\\[0.5em]
      \(m=1\) & & \(  0.514\) & \( -0.257\) & \( -0.117\) \\
      & & \( -0.315\) & \(  0.003\) & \( -0.024\) \\[0.5em]
      \(m=2\) & & \(  0.574\) & \(  0.661\) & \(  0.217\) \\
      & & \(  1.340\) & \( -0.428\) & \( -0.081\) \\[0.5em]
      \(m=3\) & & \( -1.046\) & \( -0.571\) & \( -0.184\) \\
      & & \( -1.217\) & \(  0.382\) & \(  0.045\)
      \end{tabular}
    \end{tabular}
  \end{tabular}
  \caption{Fourier coefficients used to produce the beat patterns in figures~\ref{fig:squirmer}a and \ref{fig:discreteCiliaBeats}, via equation~\eqref{eq:fourierbeat}. These coefficients were reconstructed via a least-squares procedure from the figures in \cite{blake1972model}.}\label{tab:fourier}
\end{table}

\section{Genesis of the `\(Q\)-formula'}\label{app:qformula}
\begin{figure}[h]
  \centering
  \includegraphics[angle=270,width=0.85\textwidth]{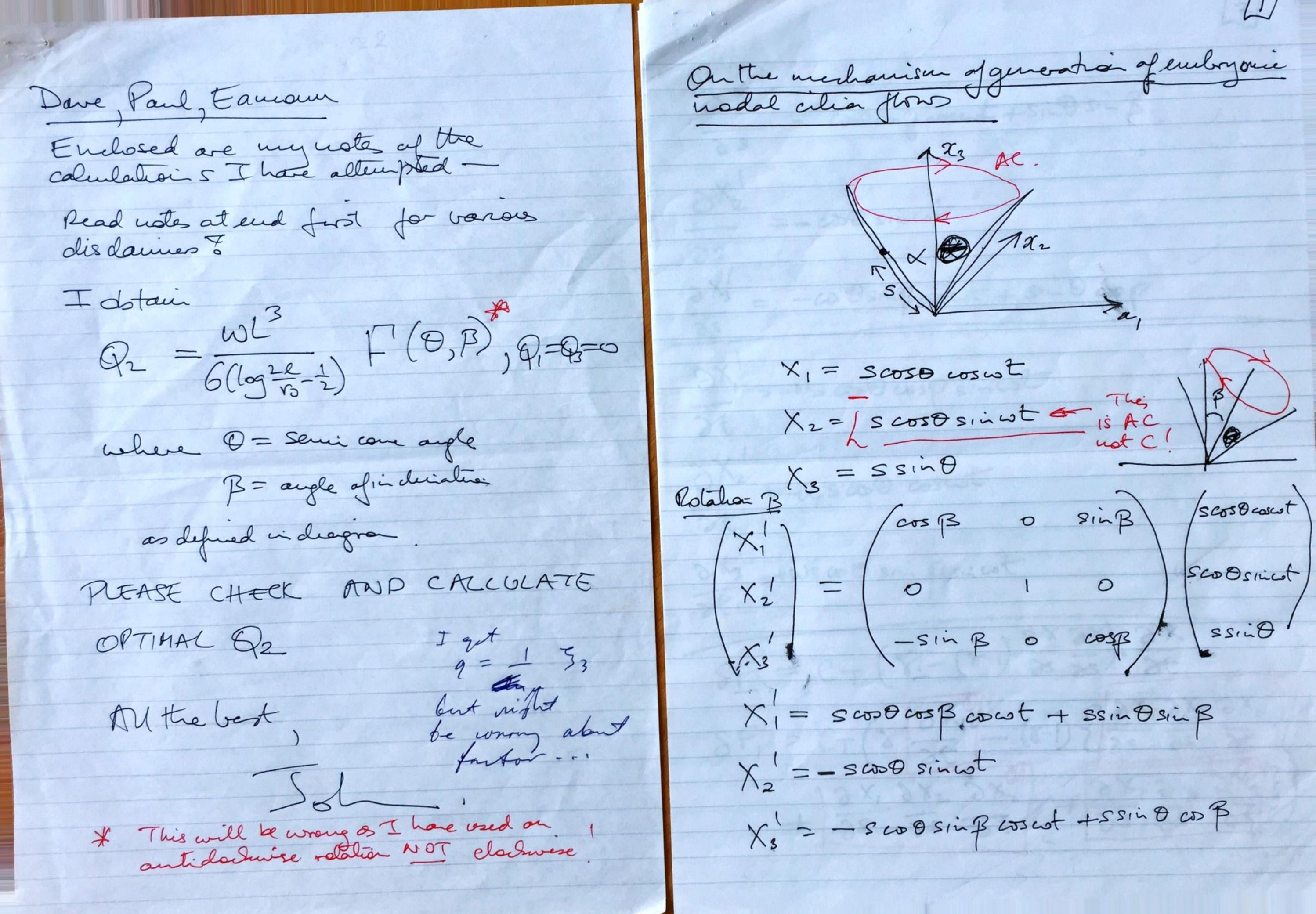}
  \caption{John's initial work on the `\(Q\)-formula' for embryonic nodal cilia in 2006, communicated by hand-written letter while recuperating in 2006 (the other individuals addressed are Paul Wakeley, then a PhD student with John, and colleague Eamonn Gaffney).}\label{fig:jrbnotes}
\end{figure}

While our simulations of particle tracks produced by tilted whirling cilia had been incorporated into ref.~\cite{smith2007discrete}, which was to be submitted in mid-2006, John was keen also to submit a shorter article addressed at a wider community than mathematical biologists. In Spring 2006, while we were producing this draft, John informed us that he was ill, and would immediately have to take sick leave. Fortunately John recovered and returned to work (as Head of School) just over a year later. Characteristically, John was not able to disconnect from scientific activity during his period of absence. Although he was often not able to meet his students and colleagues in person, he communicated through regular hand-written letters -- an example is shown in figure~\ref{fig:jrbnotes}, which shows the opening pages of John's first draft of modelling flow generation by a tilted whirling cilium, resulting (after a couple of iterations) in the `\(Q\)-formula' of equation~\eqref{eq:qformula}.

\end{appendices}


\end{document}